\documentclass[3p]{elsarticle}

\usepackage{lineno,hyperref}
\modulolinenumbers[5]

\journal{Reliability Engineering \& System Safety}

\usepackage{amsmath}
\usepackage{epsfig}
\usepackage{graphicx}
\usepackage{amsfonts}
\usepackage{relsize}
\usepackage{amsthm}

\usepackage{esvect}
\usepackage{bbm}
\usepackage{xcolor}
\usepackage{chemist}

\usepackage{algorithm}
\usepackage{algorithmic}
\usepackage[version=4]{mhchem}

\newcommand{\be}{\begin{equation}}
	\newcommand{\ee}{\end{equation}}
\newcommand{\bea}{\begin{eqnarray}}
	\newcommand{\eea}{\end{eqnarray}}

\usepackage{cancel}

\newtheorem{theorem}{Theorem}
\newtheorem{definition}{Definition}
\newtheorem{example}{Example}
\newtheorem{lemma}{Lemma}

\usepackage{caption}
\usepackage{setspace}
\begin{document}

\begin{frontmatter}

\title{Kernel-based Global Sensitivity Analysis Obtained from a Single Data Set}

\author[1]{John Barr\corref{cor1}}
\ead{jbarr@princeton.edu}
\cortext[cor1]{Corresponding author}
\author[1]{Herschel Rabitz}
\address[1]{Department of Chemistry, Princeton University, Princeton, NJ, 085044\vspace{-1cm}}

\begin{abstract}
Results from global sensitivity analysis (GSA) often guide the understanding of complicated input-output systems. Kernel-based GSA methods have recently been proposed for their capability of treating a broad scope of complex systems.  In this paper we develop a new set of kernel GSA tools when only a single set of input-output data is available.  Three key advances are made: (1) A new numerical estimator is proposed that demonstrates an empirical improvement over previous procedures. (2) A computational method for generating inner statistical functions from a single data set is presented. (3) A theoretical extension is made to define conditional sensitivity indices, which reveal the degree that the inputs carry shared information about the output when inherent input-input correlations are present.  Utilizing these conditional sensitivity indices, a decomposition is derived for the output uncertainty based on what is called the optimal learning sequence of the input variables, which remains consistent when correlations exist between the input variables.  While these advances cover a range of GSA subjects, a common single data set numerical solution is provided by a technique known as the conditional mean embedding of distributions.  The new methodology is implemented on benchmark systems to demonstrate the provided insights.
\end{abstract}


\end{frontmatter}

%\linenumbers

\section{Introduction}
Global sensitivity analysis (GSA) is frequently used to analyze how the uncertainty or variations of the input parameters influence the output of a computational model or experimental setup.  GSA has been of long standing interest to the discipline of reliability engineering and safety assessment \cite{Homma1996,Helton2006,Marrel2021}.  Pioneering works began by establishing variance-based sensitivity analysis, with the focus of developing procedures to estimate and interpret so-called Sobol indices \cite{Sobol1993,Homma1996,Rabitz1999,Sobol2003}.  Key advances to GSA include the introduction of moment-independent sensitivity analysis (MISA) \cite{Borgonovo2007}, multivariate-output aggregation \cite{Campbell2006,Lamboni2011,Gamboa2014}, goal-oriented GSA \cite{Fort2016}, and frameworks for systems with input-input correlations \cite{Li2010,Owen2017}.
In particular, Shapley effects have been of growing interest as a method when dependencies exist amongst the inputs \cite{Owen2014,Song2016,Owen2017,Iooss2019,Rabitti2019,Broto2020,Plischke2021}.

Recently, kernel-based procedures have been proposed for GSA, which have distinct advantages over many traditional approaches \cite{Barr2022,Daveiga2015,Daveiga2021}.  First, kernel GSA defines sensitivity measures for arbitrary types of input-output domains, giving methodologies that are valid for dealing with different types of data.  This has been demonstrated on systems with categorical, stochastic, time-series, functional and multivariate data \cite{Barr2022,Daveiga2021}. Second, the calculation of kernel sensitivity indices relies on  a technique known as the kernel embedding of distributions, which has been established to converge independent of the dimensionality of the output data \cite{Gretton2012}.  Thus, the corresponding GSA methods remain computationally feasible for systems with high-dimensional outputs \cite{Gretton2012,Barr2022}.  Third, kernel GSA is goal-oriented, with the underlying potential of tuning the analysis to the desired goals.  Specifically, the kernel-based GSA provides a broad class of measures that are specified by a choice of the kernel functions.  By utilizing different kernels one can determine how the input parameters influence different \textit{features} of the output \textit{distribution}.  There are two known strategies for kernel-based GSA.  One approach makes use of the \textit{maximum mean discrepancy} (MMD) as a distance metric between the unconditional and conditional output distributions \cite{Barr2022,Daveiga2021}.  Another technique relies on the Hilbert-Schmidt independence criterion (HSIC) as a distance metric between the joint input-output distribution and the product of their marginals \cite{Daveiga2015,Daveiga2021}.  Recent studies have compared these two approaches and discussed the different advantages and insights they can provide \cite{Barr2022,Daveiga2021}.

This work, focuses on the MMD method when only a single set of input-output data is available for analysis, assuming an adequate sample size; this issue will be explored in an illustration in Section \ref{sec:affine}.
In particular, many algorithms to perform kernel-based sensitivity analysis, such as the double-loop or pick-freeze procedures \cite{Barr2022,Daveiga2021}, assume that the  simulator is in a loop so that the input-output system can be conditionally re-sampled many times.  The latter assumption may be restrictive and often not viable for computationally expensive models or data obtained from experiments where new runs are called for.  We propose a new routine derived from the conditional mean embedding (CME) of distributions developed in \cite{Song2009,Grunewalder2012b,Park2020}.  This procedure has two notable improvements: first, new numerical estimators are derived that can calculate sensitivity indices from a single set of data.  Second, the estimation of CMEs allow for generating inner statistical functions, which can help visualize how the inputs marginally influence the entire output distribution. Additionally, a new theoretical extension of kernel GSA is derived to define conditional sensitivity indices.  In particular, a total output uncertainty decomposition, referred to as the optimal learning sequence (OLS), is proposed. This OLS decomposition utilizes conditional indices and remains valid in the presence of input correlations, unlike traditional ANOVA decompositions.

This paper is organized as follows.  In Section \ref{sec:kernelgsa} we review the necessary formalisms for kernel-based GSA and CMEs.  From here the new numerical techniques and the OLS decomposition are introduced.  In Section \ref{sec:compmeth} other comparative methods are discussed.  The emphasis of this section is on the nearest neighbor estimators, the kernel-ANOVA decomposition, and kernel Shapley effects proposed in \cite{Daveiga2021}.  Examples are presented on illustrative systems in Section \ref{sec:casestudies} to show the insights gained by these procedures.  Finally, conclusions are presented in Section \ref{sec:conc}.

\section{Kernel-based GSA: A review and advances}\label{sec:kernelgsa}
The theory on embedding distributions into reproducing kernel Hilbert spaces (RKHSs) and the relevant notation for GSA are briefly described in this section.  A more detailed treatment can be found in \cite{Barr2022}.  Consider a system with an input-output map $g: \mathcal{X} \rightarrow \mathcal{Y}, \mathbf{Y} = g(\mathbf{X})$,
with $\mathbf{X} \in \mathcal{X}$ and $\mathbf{Y} \in \mathcal{Y}$.  Furthermore, the input space can be represented as a product space (${\mathcal{X} = \mathcal{X}_1 \times ... \times \mathcal{X}_n}$) and the output space is an arbitrary set $\mathcal{Y}$. Note, this does not restrict the input probability measure to be a product measure, but simply enforces the inputs can be separately recorded.
In the GSA setting the input domain is assumed to be stochastic and described by the probability space $(\mathcal{X}, \mathcal{B}(\mathcal{X}), f_{\mathbf{X}})$.  In an observational/experimental setting the input distribution is innately based upon the physical phenomena and data sampling procedures.  In simulations the input distribution must be assigned by the practitioner, often guided by experimental results or expert opinion. 

Given $g$ and $(\mathcal{X}, \mathcal{B}(\mathcal{X}), f_{\mathbf{X}})$, an output probability space $(\mathcal{Y}, \mathcal{B}(\mathcal{Y}), f_{\mathbf{Y}})$ can naturally be defined.  The system's inputs and outputs are consequently understood as random variables, $\mathbf{X} \sim  f_{\mathbf{X}}$ and $\mathbf{Y} \sim  f_{\mathbf{Y}}$.  Let $(\mathbf{x},\mathbf{y})$ correspond to a realization of the random variables $(\mathbf{X},\mathbf{Y})$.
We also denote $\mathbf{X}_R$, where $R \subseteq \{1,2,...,n\}$, to specify the subset of inputs, $\mathbf{X}_R \subseteq \mathbf{X}$, that are being analyzed for their influence on the output distribution.  For shorthand, let $X_{i}$ represent $R = \{i\}$ and $\mathbf{X}_{\underline{n}}$ be $R =  \{1,2,...,n\}$.  The order refers to the number of inputs in the set $\mathbf{X}_R$, $|R|$.
Lastly, let $f_{\mathbf{X}_R}$ symbolize the marginal distribution of $\mathbf{X}_R$ and $f_{\mathbf{Y}|\mathbf{X}_R}$ be the conditional distribution of the output given $\mathbf{X}_R$, assuming a regular version.  The mean embedding of a distribution into a RKHS is defined as follows. 

\begin{definition}[RKHS and Kernel Mean Embedding]\label{def:RKHS}  Let $\mathcal{H}_k$ be a Hilbert space of $\mathbb{R}$-valued functions over the output domain  $\mathcal{Y}$.  
$\mathcal{H}_k$ is a RKHS with respect to $\mathcal{Y}$, if there exists a kernel function $k(\cdot,\mathbf{y}) \in \mathcal{H}_k, \forall \mathbf{y} \in \mathcal{Y}$, such that $f(\mathbf{y}) = \langle f(\cdot),k(\cdot,\mathbf{y})\rangle_{\mathcal{H}_k}$,  $\forall f(\cdot)  \in \mathcal{H}_k$ and $\forall \mathbf{y} \in \mathcal{Y}.$ Formally, $k: \mathcal{Y} \times \mathcal{Y} \rightarrow \mathbb{R}$ is referred to as the reproducing kernel of $\mathcal{H}_k$.	Let $\mathcal{M}_{\mathcal{Y}}$ represent the set of Borel probability measures over $\mathcal{Y}$.  The kernel mean embedding is a map of $\mathcal{M}_{\mathcal{Y}} \rightarrow \mathcal{H}_k$ defined as the expectation of the kernel,
	\begin{equation} \label{eq:kme}
		\mu_\mathbb{P} := \mathbb{E}_{\mathbf{Y} \sim \mathbb{P}}[k(\cdot,\mathbf{Y})],\hspace{0.1cm} \mathbb{P} \in \mathcal{M}_{\mathcal{Y}}.
	\end{equation}
\end{definition}

The broad applicability of RKHS theory is derived from the Moore-Aronszjan theorem, which states that if $k$ is a symmetric positive definite kernel on $\mathcal{Y}$ then it is the reproducing kernel of a unique (up to an isometry) RKHS, $\mathcal{H}_k$ \cite{Aronszajn1950}.  Crucially, this means that simply defining a symmetric positive-definite kernel function over $\mathcal{Y}$ is sufficient to apply RKHS theory, with the only restriction imposed being that $\mathcal{Y}$ is non-empty.  The MMD between two distributions can be defined as the difference of their mean embeddings.
\begin{definition}[Maximum Mean Discrepancy]\label{def:mmd} Let $\mathcal{H}_k$ be a RKHS with associated kernel, $k$, and ${\mathbb{P}_1,\mathbb{P}_2 \in \mathcal{M}_\mathcal{Y}}$.  
Given $\mathbf{Y}_1,\mathbf{Y}_1' \sim \mathbb{P}_1$ and $\mathbf{Y}_2,\mathbf{Y}_2' \sim \mathbb{P}_2$, suppose that \hspace{0.1cm} $\mathbb{E}_{\mathbf{Y}_1}\Big[\sqrt{k(\mathbf{Y}_1,\mathbf{Y}_1)}\Big] < \infty$ and \hspace{0.1cm} ${\mathbb{E}_{\mathbf{Y}_2}\Big[\sqrt{k(\mathbf{Y}_2,\mathbf{Y}_2)}\Big] < \infty}$.   If the preceding assumptions are met then $\mathbb{E}_{\mathbf{Y}_1}[k(\cdot,\mathbf{Y}_1)] = \mu_{\mathbf{Y}_1} \in \mathcal{H}_k$, $\mathbb{E}_{\mathbf{Y}_2}[k(\cdot,\mathbf{Y}_2)] = \mu_{\mathbf{Y}_2} \in \mathcal{H}_k$, and
\begin{equation}\label{eq:MMDef1}
	\text{\emph{MMD}}[\mathbb{P}_1,\mathbb{P}_2,k] = ||\mu_{\mathbf{Y}_1} -\mu_{\mathbf{Y}_2}||_{\mathcal{H}_k}.
\end{equation}
The squared MMD can be expressed in terms of the expectation of kernel functions \cite{Gretton2007},
\begin{equation} \label{eq:MMDef2}
	\text{\emph{MMD}}^2[\mathbb{P},\mathbb{Q},k]= \mathbb{E}_{\mathbf{Y}_1,\mathbf{Y}_1'}[k(\mathbf{Y}_1,\mathbf{Y}_1')]+ \mathbb{E}_{\mathbf{Y}_2,\mathbf{Y}_2'}[k(\mathbf{Y}_2,\mathbf{Y}_2')] - 2\mathbb{E}_{\mathbf{Y}_1,\mathbf{Y}_2}[k(\mathbf{Y}_1,\mathbf{Y}_2)]
\end{equation}
\end{definition}
Kernels that define Eq. \ref{eq:kme} to be injective have the special property of being \textit{characteristic}.  In applications, it is often critical to employ characteristic kernels so that probability measures are uniquely distinguished in $\mathcal{H}_k$ \cite{Gretton2012}.  The utility of characteristic kernels is given by the following theorem \cite{Gretton2007}.
\begin{theorem}\label{theor:mmd}  If the same assumptions as Definition \ref{def:mmd} are met, and $k$ is characteristic with respect to $\mathcal{Y}$, then the $\text{\emph{MMD}}[\mathbb{P}_1,\mathbb{P}_2,k] = 0$ if and only if $\mathbb{P}_1 = \mathbb{P}_2$.
\end{theorem}\noindent
Ultimately, Theorem \ref{theor:mmd} concerns how moment-independent measures are encapsulated into the of class kernel-based GSA procedures \cite{Barr2022,Daveiga2021}.  With the notation formally introduced we will now discuss kernel GSA.
\begin{definition}[$\beta^k$-Indicator]\label{def:betafirst} Let $k$ be a symmetric positive definite kernel on $\mathcal{Y}$, with corresponding RKHS $\mathcal{H}_k$.  The following $\beta^k$-indicator,
	\begin{equation}\label{eq:betadef1}
		\beta_{R}^k := \frac{\mathbb{E}_{\mathbf{X}_R}\big[\text{\emph{MMD}}^2[f_{\mathbf{Y}},f_{\mathbf{Y}|\mathbf{X}_R},k]\big]}{\mathbb{E}_{\mathbf{X}_{\underline{n}}}\big[\text{\emph{MMD}}^2[f_{\mathbf{Y}},f_{\mathbf{Y}|\mathbf{X}_{\underline{n}}},k]\big]},
	\end{equation}
	yields an average normalized change in the output distribution induced by determining $\mathbf{X}_R$.  Equation \ref{eq:betadef1} has the following equivalent forms \cite{Barr2022},
	\begin{equation}\label{eq:betaVdef}
		\beta_{R}^k = \frac{\mathbb{E}_{\mathbf{X}_R}\big[\text{\emph{MMD}}^2[f_{\mathbf{Y}},f_{\mathbf{Y}|\mathbf{X}_R},k]\big]}{\mathbb{E}_{\mathbf{y}\sim f_{\mathbf{Y}}}[k(\mathbf{y},\mathbf{y})] - \mathbb{E}_{\mathbf{y},\mathbf{y}'\sim f_{\mathbf{Y}}}[k(\mathbf{y},\mathbf{y}')]},
	\end{equation}
and 
\begin{equation}\label{eq:betaUdef}
	\beta_{R}^k = \frac{\mathbb{E}_{\mathbf{X}_R}\big[ \mathbb{E}_{\mathbf{y},\mathbf{y}' \sim f_{\mathbf{Y}|\mathbf{X}_R}}[k(\mathbf{y},\mathbf{y}')|\mathbf{X}_R]\big] - \mathbb{E}_{\mathbf{y},\mathbf{y}'\sim f_{\mathbf{Y}}}[k(\mathbf{y},\mathbf{y}')]}{\mathbb{E}_{\mathbf{y}\sim f_{\mathbf{Y}}}[k(\mathbf{y},\mathbf{y})] - \mathbb{E}_{\mathbf{y},\mathbf{y}'\sim f_{\mathbf{Y}}}[k(\mathbf{y},\mathbf{y}')]}.
\end{equation}
\end{definition}

The logic of this metric follows the generalized rationale of Borgonovo et al. \cite{Borgonovo2016b}.  It begins with the base case of all input parameters free to vary ($\mathbf{X}_{\underline{n}} \sim f_{\mathbf{X}}$), so that the output follows the unconditional distribution, $f_{\mathbf{Y}}$.  Next, suppose that the uncertainty from an input or combination of input sources, say $\mathbf{X}_R$, can be eliminated from analysis by determining $\mathbf{X}_R = \mathbf{x}_R$. In this scenario the outputs would now follow the conditional distribution $f_{\mathbf{Y} | \mathbf{X}_R = \mathbf{x}_R}$.  This is often simplified to $f_{\mathbf{Y} | \mathbf{X}_R}$ as $\mathbf{X}_R = \mathbf{x}_R$ is implicitly assumed. The change in the output distribution, upon reducing the source of uncertainty, can be quantitatively assessed through a divergence measure referred to as the inner statistic function (ISF).   Based on Definition \ref{def:betafirst}, there are two ISFs that can be used to determine a $\beta_R^k$ value.  The first is,
\begin{equation}\label{eq:ISFDdef}
	\gamma_R^D(\mathbf{x}_R) = \text{MMD}^2[f_{\mathbf{Y}},f_{\mathbf{Y}|\mathbf{X}_R},k],
\end{equation}
from Eq. \ref{eq:betaVdef} and the second is,
\begin{equation}\label{eq:ISFNdef}
	\gamma_R^N(\mathbf{x}_R) = ||\mu_{\mathbf{Y}|\mathbf{X}_R}||_{\mathcal{H}_k}^2 = \mathbb{E}_{\mathbf{y},\mathbf{y}' \sim f_{\mathbf{Y}|\mathbf{X}_R}}[k(\mathbf{y},\mathbf{y}')|\mathbf{X}_R].
\end{equation}
for Eq. \ref{eq:betaUdef}.  The superscript D indicates the ISF that utilizes the MMD \textit{distance} metric while the superscript N denotes the ISF based on the \textit{norm} in $\mathcal{H}_k$.  The information is reduced to a single index, $\beta_R^k$, by averaging the ISFs with respect to the input distribution.  In summary, a $\beta_R^k$ index assesses the average change in the output distribution (measured in the feature space $\mathcal{H}_k$) caused by learning input(s) $\mathbf{X}_R$.  The denominator given in Definition \ref{def:betafirst} is a normalization coefficient that provides interpretable bounds for $\beta_R^k$. 

Note that every symmetric positive definite kernel will define a GSA metric given by $\beta_R^k$.  Changing the kernel will result in altering the \textit{feature} of the \textit{output distribution} that is emphasized.  The fact that changing the kernel alters the sensitivity measure is a significant matter which is expanded upon in \cite{Barr2022}.  The $\beta^k$-indicator is normalized, meaning $0 \leq \beta^k_R \leq 1$ and $\beta_{\underline{n}}^k = 1$.  Further, the $\beta^k$-indicator is strictly increasing with respect to the amount of inputs accounted for, if $\mathbf{X}_S \subseteq \mathbf{X}_R$ then $\beta_{S}^k \leq \beta_{R}^k$.  These properties hold for any symmetric positive-definite kernel, even if $k$ is not characteristic.  However, when $k$ is characteristic the $\beta^k$-indicator will be moment-independent.  Thus, it not strictly necessary for $k$ to be characteristic to define a valid sensitivity measure through Definition \ref{def:betafirst}. The work of \cite{Barr2022} highlights how meaningful $\beta^k$-indicators, such as variance-based routines, arise from non-characteristic kernels.

There has been extensive work to understand the various properties of kernels and their relationship between one another \cite{Gretton2012,Sriperumbudur2010,Sriperumbudur2011}.  Additionally, there are many known kernels functions, with works such as \cite{Muandet2017}, providing lists of kernels and their well-established properties.  Heuristics are available that can be employed to choose kernels and kernel hyperparameters \cite{Barr2022,Daveiga2021}.  For example, if the goal is to use a moment-independent GSA measure with $\mathcal{Y} \subseteq \mathbb{R}^m$, then the Gaussian radial basis function (RBF) is a typical choice for a characteristic kernel,
\begin{equation}
	k(\mathbf{y},\mathbf{y}')=\text{exp}\big(-\frac{||\mathbf{y}-\mathbf{y}'||_2^2}{2\sigma^2}\big);\, \sigma >0.
\end{equation}
Selecting the bandwidth hyperparameter, $\sigma$, as either the median distance of the output data or the square root of the sum of variances are useful choices often used in practice \cite{Barr2022,Daveiga2021}.  

An optimal kernel choice can be a consideration, either to compare different families of kernels and/or choose kernel hyperparameters, although a prescriptive procedure for \textit{a priori} kernel selection is presently not available for GSA.   There have been proposed solutions for an optimal kernel choice in other statistical contexts where the embedding of distributions can be applied.  For example, in \cite{Gretton2012b} the focus is on employing kernel embeddings for the two-sample test, where the kernel is chosen to maximize the power of the test.  It has been postulated in works such as \cite{Barr2022,Daveiga2021} that a key element to optimal kernel selection depends on developing ``central limit theorems" for the estimators of Definition \ref{def:betafirst}.  In essence, if the data can be thought of as random variables (e.g. due to the nature of their sampling), then any numerical indices that are functions of the data would also be random variables.  Understanding the features of these approximations could ultimately lead to procedures that maximize the $\beta^k$-indicators ability to detect input-output dependencies.  However, the focus of this work is to first establish a new data-driven methodology and therefore such analyses fall outside of the purview of this paper.  In the case studies presented in Section \ref{sec:casestudies}, we will utilize the heuristics of \cite{Barr2022} to choose the output kernels.

\subsection{Application of Conditional Mean Embeddings}\label{sec:CME}
This section describes the CME, followed by proposing a new algorithm for determining ISFs and calculating $\beta_R^k$ indices.
The CME was first proposed in \cite{Song2009} and was defined by composing cross-covariance operators.  The theoretical consistency of this definition relied on some overly strict assumptions, which are often violated and not guaranteed \cite{Fukumizu2013,Klebanov2020}.  Park et al. recently developed a measure theoretic approach to defining the CME that lifts these assumptions \cite{Park2020} and it is used here.
\begin{definition}[Conditional Mean Embedding]\label{def:cmeog} Let $k: \mathcal{Y} \times \mathcal{Y} \rightarrow \mathbb{R}$ be symmetric positive-definite.  Assuming $\mathbb{E}_{\mathbf{y} \sim f_{\mathbf{Y}|\mathbf{X}_R}}[k(\mathbf{y},\mathbf{y})|\mathbf{X}_R=\mathbf{x}_R] < \infty$ almost surely, the conditional mean embedding is defined as,
	\begin{equation}
		\mu_{\mathbf{Y}|\mathbf{X}_R}(\cdot|\mathbf{x}_R) := \mathbb{E}_{\mathbf{y} \sim f_{\mathbf{Y}|\mathbf{X}_R}}[k(\cdot,\mathbf{y})|\mathbf{X}_R=\mathbf{x}_R]
	\end{equation}
\end{definition}
This definition extends from the unconditional mean embedding given by Eq. \ref{eq:kme}.  However, unlike the unconditional embedding, the CME does not represent a single element in $\mathcal{H}_k$, but instead covers a wide family of points designated by $\mathbf{x}_R$.  Park et al. show that the CME is a deterministic function that is measurable with respect $\mathcal{X}_R$ and the Borel $\sigma$-algebra of $\mathcal{H}_k$, $\mathcal{B}(\mathcal{H}_k)$ \cite{Park2020}.  This makes the determination of $\mu_{\mathbf{Y}|\mathbf{X}_R}(\mathbf{x}_R)$ from data a vector-valued regression problem with the input space $\mathcal{X}_R$ being used to predict elements of the RKHS, $\mathcal{H}_k$. Following \cite{Grunewalder2012b}, let $\mathcal{G}$ be a vector-valued RKHS of functions, with kernel $G$.  For simplicity, choose $G(\mathbf{x}_R,\mathbf{x}_R') = l_R(\mathbf{x}_R,\mathbf{x}_R')\mathbf{I}$, where $\mathbf{I}$ is the identity operator and the function ${l_R: \mathcal{X}_R \times \mathcal{X}_R \rightarrow \mathbb{R}}$ is a real-scalar valued kernel over the input domain $\mathbf{X}_R$.  Given a set of joint input-output data, $\mathcal{D}^N = \big\{(\mathbf{x}_R^{(1)},\mathbf{y}^{(1)}), (\mathbf{x}_R^{(2)},\mathbf{y}^{(2)}),...,(\mathbf{x}_R^{(N)},\mathbf{y}^{(N)})\big\}$, a natural data driven loss functional can be expressed as \cite{Park2020},
\begin{equation}\label{eq:cmeloss2}
	\mathcal{E}[F] = \frac{1}{N}\sum_{i=1}^{N}||k(\cdot,\mathbf{y}^{(i)}) - F(\mathbf{x}_R^{(i)})||^2_{\mathcal{H}_k} + \lambda||F||_{\mathcal{G}}^2, \text{ } F \in \mathcal{G},
\end{equation}
where $\lambda>0$ is the regularization parameter included to keep the optimization problem well-posed.  Note that $\mathcal{E}$ is not the only option for an error functional, with works such as \cite{Laforgue2020} exploring other loss functionals and \cite{Kadri2016} considering vector-valued kernels outside of $G=l_R(\mathbf{x}_R,\mathbf{x}_R')\mathbf{I}$.

The advantage to choosing $\mathcal{E}$ as defined in Eq. \ref{eq:cmeloss2} is that the solution has a closed form (proven in \cite{Grunewalder2012b}),
\begin{equation}\begin{split}\label{eq:CMEF}
	\hat{\mu}_{\mathbf{Y}|\mathbf{X}_R}(\mathbf{x}_R) &:= \text{argmin}_{F}\{\mathcal{E}[F] \} = \mathbf{\Gamma}_R^T(\mathbf{x}_R)\mathbf{W}\mathbf{\Phi}(\cdot)\\
	\mathbf{W} &= (\mathbf{L}_R + \lambda\mathbf{I}_N)^{-1},
\end{split}\end{equation}
where $\mathbf{I}_N$ is the $N \times N$ identity matrix, $\mathbf{\Gamma}_R$ and $\mathbf{\Phi}$ are the feature vectors of the inputs and outputs, respectively, and $\mathbf{L}_R$ is the Gram matrix of the input data,
\begin{equation}\begin{split}
		\mathbf{\Phi}(\cdot) &= (k(\cdot,\mathbf{y}^{(1)}),k(\cdot,\mathbf{y}^{(2)}),...,k(\cdot,\mathbf{y}^{(N)})))^T\\
		\mathbf{\Gamma}(\cdot)&=(l_R(\cdot,\mathbf{x}_R^{(1)}),l_R(\cdot,\mathbf{x}_R^{(2)}),...,l_R(\cdot,\mathbf{x}_R^{(N)}))^T.\\
		[\mathbf{L}_R]_{ij} &= l_R(\mathbf{x}_R^{(i)},\mathbf{x}_R^{(j)})
\end{split}\end{equation}  
The consistency of Eq. \ref{eq:CMEF} is given by \cite{Park2020} under the assumptions that the kernel $l_R$ is bounded and universal in $\mathbf{X}_R$.  These criteria restrict the input kernel to be universal in order to guarantee convergence, but these assumptions are non-restrictive since there are well-known kernels with the universal property that can be used in practice, such as the Gaussian RBF.  Additionally, working in the regression setting allows for cross-validation, which facilitates choosing input kernel hyperparameters and comparing different kernels.  There are a wide-choice of data-driven error functions that could be used to test and validate a CME estimate; we utilize a K-fold cross-validation proposed by Gr\"{u}new\"{a}lder et al. \cite{Grunewalder2012b}.

With the CME process established we now propose a new set of kernel GSA tools, beginning with the ISFs.  To account for normalization, the U-statistics can be used as approximations for the terms in the denominator  \cite{Serfling2009},
\begin{equation}\begin{split}\label{eq:shorthand}
		\mathbb{E}_{\mathbf{y}\sim f_{\mathbf{Y}}}[k(\mathbf{y},\mathbf{y})] \approx\text{ }  \hat{C}_{\mathbf{Y}} & := \frac{1}{N}\sum_{i=1}^N k(\mathbf{y}^{(i)},\mathbf{y}^{(i)}) = \frac{\text{Tr}[\mathbf{K}]}{N}\\
		\mathbb{E}_{\mathbf{y},\mathbf{y}'\sim f_{\mathbf{Y}}}[k(\mathbf{y},\mathbf{y}')] \approx  \hat{C}_{\mathbf{Y},\mathbf{Y}}  := &\frac{1}{N(N-1)}  \sum_{\substack{i,j=1 \\ i\neq j}}^{N,N} k(\mathbf{y}^{(i)},\mathbf{y}^{(j)}) = \frac{\mathbf{1}_N^T(\mathbf{K} - \mathbf{I}_N)\mathbf{1}_N}{N(N-1)},
\end{split}\end{equation}
where $\mathbf{K}$ is the Gram matrix of the output data, $[\mathbf{K}]_{ij} = k(\mathbf{y}^{(i)},\mathbf{y}^{(j)})$, and $\mathbf{1}_N \in \mathbb{R}^N$ is a $N\times1$ vector with each element equal to 1.  The terms $\hat{C}_{\mathbf{Y}}$ and $\hat{C}_{\mathbf{Y},\mathbf{Y}}$ are introduced as shorthand notation. Since $\hat{C}_\mathbf{Y}$ and $\hat{C}_{\mathbf{Y},\mathbf{Y}}$ are constants independent of $\mathbf{X}_R$ that will converge to $\mathbb{E}_{\mathbf{y}\sim f_{\mathbf{Y}}}[k(\mathbf{y},\mathbf{y})]$ and $\mathbb{E}_{\mathbf{y},\mathbf{y}'\sim f_{\mathbf{Y}}}[k(\mathbf{y},\mathbf{y}')]$ respectively, they can be exchanged with the input expectation operator in Eqs. \ref{eq:betaVdef} and \ref{eq:betaUdef} to generate an ISF appropriately scaled with reference to the unconditional output distribution.

\begin{definition}[Inner Statistical Functions]\label{def:ISF} The following function of $\mathbf{x}_R$,
	\begin{equation}\vspace{-0.1cm}
		\overline{\gamma}_{R}^N(\mathbf{x}_R) = \frac{||\mu_{\mathbf{Y}|\mathbf{X}_R}(\mathbf{x}_R)||^2_{\mathcal{H}_k}- \mathbb{E}_{\mathbf{y},\mathbf{y}'\sim f_{\mathbf{Y}}}[k(\mathbf{y},\mathbf{y}')]}{\mathbb{E}_{\mathbf{y}\sim f_{\mathbf{Y}}}[k(\mathbf{y},\mathbf{y})] - \mathbb{E}_{\mathbf{y},\mathbf{y}'\sim f_{\mathbf{Y}}}[k(\mathbf{y},\mathbf{y}')]},\\
	\end{equation}\vspace{-0.1cm}
	describes how the norm of the conditional mean embedding in $\mathcal{H}_k$ changes with respect to $\mathbf{x}_R$.  This can be approximated as,\vspace{-0.1cm}
	\begin{equation}\label{eq:Nest}
		\hat{\gamma}_{R}^N(\mathbf{x}_R) :=  \frac{||\hat{\mu}_{\mathbf{Y}|\mathbf{X}_R}(\mathbf{x}_R)||^2_{\mathcal{H}_k}-\hat{C}_{\mathbf{Y},\mathbf{Y}}}{\hat{C}_{\mathbf{Y}} - \hat{C}_{\mathbf{Y},\mathbf{Y}}} = \frac{\mathbf{\Gamma}_R^T(\mathbf{x}_R)\mathbf{W}\mathbf{K}\mathbf{W}\mathbf{\Gamma}_R(\mathbf{x}_R)-\hat{C}_{\mathbf{Y},\mathbf{Y}}}{\hat{C}_{\mathbf{Y}} - \hat{C}_{\mathbf{Y},\mathbf{Y}}} \\
	\end{equation}\vspace{-0.1cm}
	Additionally, the function,\vspace{-0.1cm}
	\begin{equation}
		\overline{\gamma}_{R}^D(\mathbf{x}_R) = \frac{||\mu_{\mathbf{Y}|\mathbf{X}_R}(\mathbf{x}_R)-\mu_{\mathbf{Y}}||^2_{\mathcal{H}_k}}{\mathbb{E}_{\mathbf{y}\sim f_{\mathbf{Y}}}[k(\mathbf{y},\mathbf{y})] - \mathbb{E}_{\mathbf{y},\mathbf{y}'\sim f_{\mathbf{Y}}}[k(\mathbf{y},\mathbf{y}')]},\\
	\end{equation}\vspace{-0.1cm}
	quantifies how the distance between the conditional mean embedding and the unconditional mean embedding in $\mathcal{H}_k$ changes with respect to $\mathbf{x}_R$.  This is determined as,\vspace{-0.1cm}
	\begin{equation}\begin{split}\label{eq:Dest}
			\hat{\gamma}_{R}^D(\mathbf{x}_R) &:= \frac{||\hat{\mu}_{\mathbf{Y}|\mathbf{X}_R}(\mathbf{x}_R)-\hat{\mu}_{\mathbf{Y}}||^2_{\mathcal{H}_k}}{\hat{C}_{\mathbf{Y}} - \hat{C}_{\mathbf{Y},\mathbf{Y}}}\\
			&=\frac{\mathbf{1}_N^T\mathbf{K}\mathbf{1}_N+\mathbf{\Gamma}_R^T(\mathbf{x}_R)\mathbf{W}\mathbf{K}\mathbf{W}\mathbf{\Gamma}_R(\mathbf{x}_R) - 2\cdot\mathbf{1}_N\mathbf{K}\mathbf{W}\mathbf{\Gamma}_R(\mathbf{x}_R)}{\hat{C}_{\mathbf{Y}} - \hat{C}_{\mathbf{Y},\mathbf{Y}}},
	\end{split}\end{equation}
\end{definition}

The approximations given by Eq. \ref{eq:Nest} and Eq. \ref{eq:Dest} are derived by substituting Eq. \ref{eq:CMEF} in for the CME and $\hat{\mu}_{\mathbf{Y}} = \frac{1}{N}\sum_{i=1}^{N}k(\cdot,\mathbf{y}^{(i)})$ for the unconditional mean embedding.  From here expanding the norm and applying the reproducing property is sufficient to get the closed matrix forms.  The ISFs derived from Definition \ref{def:ISF} provide a means to see how inputs, either individually with first-order functions or cooperatively with higher-order functions, influence the output \textit{distribution} based on the features defined by $k$.  The function $\overline{\gamma}_{R}^N(\mathbf{x}_R)$ will yield how the squared norm of $\hat{\mu}_{\mathbf{Y}|\mathbf{X}_R}$ in $\mathcal{H}_k$ changes with respect to $\mathbf{x}_R$, which measures the spread of the output data with respect to the features highlighted by $k$. Furthermore, $\overline{\gamma}_{R}^D(\mathbf{x}_R)$ describes the difference between the unconditional and conditional output distribution.  Note that $\beta_R^k = \mathbb{E}_{\mathbf{X}_R}[\overline{\gamma}_{R}^N(\mathbf{x}_R)] =  \mathbb{E}_{\mathbf{X}_R}[\overline{\gamma}_{R}^D(\mathbf{x}_R)]$ in Definition \ref{def:betafirst}, so Eqs. \ref{eq:Nest} and \ref{eq:Dest} enable the efficient calculation of $\beta_R^k$ indices as the empirical average of the ISFs,
\begin{equation}\label{eq:Fest}
\hat{\beta}_R^{k,L} = \frac{1}{N}\sum_{i=1}^N\hat{\gamma}^L_R(\mathbf{x}_R^{(i)}),\, L \in \{N,D\}.
\end{equation}
The consistency of the estimators extend from the consistency proofs of the CME given by \cite{Park2020} under the same assumptions of a bounded universal input kernel.  In summary, two new routines for the $\beta_R^k$ indices are given by Eq. \ref{eq:Fest}.  Additionally, a numerical determination of the ISFs, which provide different insights about how the inputs influence the output distribution, is given by Eqs. \ref{eq:Dest} and  \ref{eq:Nest}.

\subsection{Conditional Indices and Optimal Learning Sequence Decomposition}\label{sec:olo}
The objective of this section is to introduce a new decomposition of the total output uncertainty, $\beta_{\underline{n}}=1$.  To begin, a conditional kernel-based sensitivity index is introduced. 
\begin{definition}[Conditional Index]\label{def:CI} Let $\mathbf{X}_D = \mathbf{X}_S \cup \mathbf{X}_R$ and $\mathbf{X}_S \cap \mathbf{X}_R = \emptyset$ so $\mathbf{X}_S \subseteq \mathbf{X}_D \subseteq \mathbf{X}_{\underline{n}}$. A conditional $\beta^k_{R|S}$ index is determined as,
	\begin{equation}\label{eq:CI}
		\beta_{R|S}^k=\frac{\mathbb{E}_{\mathbf{X}_{\underline{n}}}\big[\text{\emph{MMD}}^2[f_{\mathbf{Y}|\mathbf{X}_D},f_{\mathbf{Y}|\mathbf{X}_S},k]\big]}{\mathbb{E}_{\mathbf{y}\sim f_{\mathbf{Y}}}[k(\mathbf{y},\mathbf{y})] - \mathbb{E}_{\mathbf{y},\mathbf{y}'\sim f_{\mathbf{Y}}}[k(\mathbf{y},\mathbf{y}')]}.
	\end{equation}
\end{definition}\noindent
These conditional indices lead to a hierarchical approach to decomposing the uncertainty based upon the following lemma [Proven in \cite{Barr2022}, Supplemental Lemma SM4.1].
\begin{lemma}\label{lem:condlem} If $\mathbf{X}_D = \mathbf{X}_S \cup \mathbf{X}_R$ and $\mathbf{X}_S \cap \mathbf{X}_R = \emptyset$ then $\beta_D^k=\beta_S^k+\beta_{R|S}^k$
\end{lemma}
Lemma \ref{lem:condlem} yields a component-wise breakdown of the output uncertainty apportioned to a set of inputs.  For example, Lemma \ref{lem:condlem} gives the following two equivalencies,  ${\beta_{(1,2)}^k = \beta_{1}^k + \beta_{2|1}^k = \beta_{2}^k + \beta_{1|2}^k.}$
The interpretation of this equation is that the uncertainty in the output distribution, which can be accounted for by simultaneously determining $X_1$ and $X_2$, is equivalent to first learning input $X_1$ and then learning $X_2$ when given $X_1$.  Parallel logic applies to the second equality.  Note that higher-order conditional indices also exist, i.e. $\beta_{(3,4)|(1,2)}$ representing the conditional information from learning inputs $X_3$ and $X_4$ when inputs $X_1$ and $X_2$ are already determined.  Further, the order in which the variables are learned in will relay different information about the output uncertainty accounted for at each step (if $\beta_{1} \neq \beta_{2}$ then $\beta_{1|2} \neq \beta_{2|1}$).

Importantly, Lemma \ref{lem:condlem} endows the $\beta^k$-indicator with the property that the information gained from learning inputs simultaneously is equivalent to determining them piece-wise.  Thus, the following total decomposition can be shown to hold (by induction):
\begin{equation}\label{eq:decomp1}
	\beta_{\underline{n}}^k = 1 = \beta_1^k + \beta_{2|1}^k + \beta_{3|(1,2)}^k + \beta_{4|(1,2,3)}^k + ... + \beta_{n|(1,2,...,n-1)}^k.
\end{equation}
The rationale behind Eq. \ref{eq:decomp1} is that the total output uncertainty ($\beta_{\underline{n}}^k = 1$) can be apportioned to each of the inputs as they are learned one-at-at-time in a particular sequence.  The specific sequence in Eq. \ref{eq:decomp1} represents learning the inputs in the same order as their assigned labels, $1,...,n$.  This decomposition is non-unique as the inputs can be learned in any particular sequence, and given $n$ input variables there would be $n!$ total different sequences.  We propose to use a specific sequence, that we refer to as the optimal learning sequence (OLS).  The premise of the OLS is that if each input variable could be learned in a prescribed sequence then it should be done so that at each step the input is chosen to maximally reduce the uncertainty in the output.
\begin{definition}[Optimal Learning Sequence Decomposition]\label{def:olo} Let $i_l,l \in \{1,...,n\}$ be input labels where,
\begin{itemize}
		\item $i_1$ is the label of the largest first-order index: $i_1=\underset{j}{\max}\big[\beta_j^k\big]$. 
		\item $i_2$ is the label of the largest conditional index given $i_1$: $i_2=\underset{j}{\max}\big[\beta_{j|i_1}^k\big]$. 
		\item $i_3$ is the label of the largest conditional index given $i_1$ and $i_2$: $i_3=\underset{j}{\max}\big[\beta_{j|(i_1,i_2)}^k\big]$.
		\item etc.
\end{itemize}
The OLS decomposition is then defined as,
\begin{equation}\label{eq:decomp2}
	 \beta_{i_1}^k + \beta_{i_2|i_1}^k + \beta_{i_3|(i_1,i_2)}^k  + ... + \beta_{i_n|(i_1,i_2,...,i_{n-1})}^k = 1.
\end{equation} 
\end{definition}

This result assumes that such a sequence is unique, or there is a single maximum index at a given step.  If a non-unique scenario were to arise, then it would be most informative to consider multiple cases and compare additional conditional indices.  For example, if $\beta_1^k = \beta_2^k$ then the most information about the system will be revealed by generating both sequences where $i_1 = 1$ and $i_1 = 2$.

There are some final theoretical implications behind Lemma \ref{lem:condlem} that are detailed here.  First, Lemma \ref{lem:condlem} does not require any additional assumptions on the kernel.  Thus, the existence of a decomposition given by Eq. \ref{eq:decomp1} immediately follows for any GSA measure that arises as a particular case of the $\beta^k$-indicator (such as variance-based methods).  Second, conditional indices can be determined exactly from the unconditional set, so there is no additional computational cost.  In general, however, the sheer number of both unconditional and conditional indices grow exponentially.  Particularly, given $n$ inputs there would be $2^n - 1$ \textit{unconditional} and $3^n - 2^{n+1}+1$ \textit{conditional} indices.  In practice one often reports a small subset of results for the input variables that are found to be important.  Following this practice, determining the OLS only requires the calculation of $n(n+1)/2$ unconditional indices for a full decomposition (assuming a unique sequence), so there is no exponential cost in reporting the influence of the input variables through this procedure.

\section{Comparative Methods}\label{sec:compmeth}
This section considers methods from the literature which mainly come from the work of \cite{Daveiga2021}.  There are two points of focus in this section, (1) numerical procedures to calculate $\beta_R^k$ and (2) theoretical tools to interpret higher-order effects in the presence of input correlations.  We remark here that \cite{Daveiga2021} does not utilize the terminology $\beta^k$-indicator, but by applying the normalization of Definition \ref{def:betafirst} the same results are obtained with the notation used in this paper.

\subsection{Numerical Procedures}\label{sec:compnum}
Previously proposed procedures for calculating $\beta_R^k$ indices have included double-loop Monte Carlo methods \cite{Barr2022,Daveiga2021}, a pick-freeze approach \cite{Daveiga2021}, and a rank-order/nearest-neighbor routine \cite{Daveiga2021}.  Importantly, the double-loop and pick-freeze techniques require the ability to draw upon multiple data sets based upon conditional sampling, which is a stringent requirement.  Specifically, if one only has access to a single set of input-output data, then the procedures can not be applied.  The nearest-neighbor approach in contrast can be employed with a single data set.  Adapting the notation of \cite{Daveiga2021,Broto2020}, let the function $j_R^{*}(i,m)$ return the index of the $m$-th nearest neighbor of input sub-sample point $\mathbf{x}_R^{i}$.  Noting $j_R^{*}(i,1)=i$, the following estimator based on utilizing a single data set is introduced as,
\begin{equation}\label{eq:NNEFull}
	\hat{C}_{\mathbf{X}_R}^{F}=\frac{1}{N}\sum_{i=1}^{N}k\big(\mathbf{y}^{(i)},\mathbf{y}^{(j_R^{*}(i,2))}\big).
\end{equation}

Eq. \ref{eq:NNEFull} can often have bias, as noted by \cite{Daveiga2021}, thus an estimator based on a random sub-sample is introduced following the work of \cite{Broto2020},
\begin{equation}\label{eq:NNESubsample}
	\hat{C}_{\mathbf{X}_R}^{S}=\frac{1}{N_A}\sum_{i=1}^{N_A}k\big(\mathbf{y}^{(s(i))},\mathbf{y}^{(j_R^{*}(s(i),2))}\big).
\end{equation}
In this situation $i \in \{1,..., N_A\}$, with $N_A \leq N$, and $s(i)$ are randomly chosen integers uniformly distributed over $\{1,...,N\}$.  The utilization of a random sub-sample is motivated by the work in \cite{Broto2020} which proposed the procedure for calculating Sobol sensitivity indices.  Da Veiga \cite{Daveiga2021} remarked that the nearest-neighbor estimate had notable bias for higher-order indices, and extended the technique of \cite{Broto2020} to reduce this bias compared to the full-sample estimator.  From Eqs. \ref{eq:NNEFull} and \ref{eq:NNESubsample}, the $\beta_R^k$ indices can be approximated as,
\begin{equation}\label{eq:NNjoint}
	\hat{\beta}_R^{k,L} = \frac{\hat{C}_{\mathbf{X}_R}^{L} - \hat{C}_{\mathbf{Y},\mathbf{Y}}}{\hat{C}_{\mathbf{Y}} - \hat{C}_{\mathbf{Y},\mathbf{Y}}}, L \in \{F,S\}
\end{equation}
where $\hat{C}_{\mathbf{Y}}$ and $\hat{C}_{\mathbf{Y},\mathbf{Y}}$ were previously given by Eq. \ref{eq:shorthand}.  The labels $F$ and $S$ are used to represent the nearest-neighbor estimates that use the full-sample and random sub-samples, respectively.

To summarize, a comparison between different numerical methodologies is presented in Table \ref{tab:methcomp}.  The total simulation cost column represents the number of input-output function calls required for a full-order sensitivity analysis, assuming data sets of size $N$ would be used to calculate the MMD with Eq.  \ref{eq:MMDef2}.  This number would be reduced for both the double-loop and pick-freeze techniques if the analysis was only directed at subset of results, such as strictly a first-order analysis.  For the double-loop approach there is an additional data sampling parameter, $N_{ext}$, that represents the number of conditional data sets generated for the empirical average of the MMD in Eq. \ref{eq:betadef1} (for more details see \cite{Barr2022}).  The ISF column of Table \ref{tab:methcomp} indicates if the numerical procedure yields additional information about the marginal influence of the inputs on the output distribution through determination of the ISFs.

\begin{table}[h]
	\begin{center}
		\caption{Comparison of Different Methods.\label{tab:methcomp}}
		\begin{tabular}{c c c}
			\hline
			Method & Total Simulation Cost & ISFs?\\
			\hline
			Double-Loop & $(2^n-1)N_{ext}N+N$ & No\\
			Pick-Freeze & $2^nN$ & No\\
			Nearest Neighbor (NN) & $N$ & No\\
			Conditional Mean Embedding (CME) & $N$ & Yes\\
			\hline
\end{tabular}\end{center}\end{table}\vspace{-0.2cm}

Table \ref{tab:methcomp} shows the significance of operating from only a single set of input-output data.   Specifically, there is an exponential cost in total function calls when a particular method requires conditionally re-sampling, which effectively restricts such strategies to input-output models with a cheap time-cost.  Out of the available procedures, only two fit the single-sample criterion, the nearest-neighbor technique of \cite{Daveiga2021} and the CME approach presented here.  In comparing these two strategies, the nearest-neighbor and CME, there are two final points to consider.  First, the CME can provide additional insights through the numerical determination of the ISFs, which is not presently available through nearest-neighbor routines.  Further, the work of \cite{Daveiga2021} empirically show that nearest-neighbor approaches, both the sub-sampling and full-sample estimators of Eq. \ref{eq:NNjoint}, often have bias.  The CME offers an improvement in this regard, with an example presented in Section \ref{sec:affine} that demonstrates a reduced bias and higher accuracy compared to the nearest-neighbor estimates.  A detailed study can be found in \cite{Deb2020}, where a kernel-method is presented for measuring the statistical association of two topological spaces, using the same measure given by Eq. \ref{eq:betaUdef}.  This work provides theorems on the asymptotic behavior of a family nearest-neighbor estimates, including insights into the inherent variance and bias of these estimators.  However, to fully compare the two methods, such central limit theorems are necessary for the CME techniques (in terms of accuracy and bias).  Such theorems could also provide insights into necessary data set sizes to ensure a sufficient accuracy.  An analytical example is presented in Section \ref{sec:affine}, which shows larger error for smaller samples, which is typical of data driven analysis.

\subsection{Kernel-ANOVA Decomposition and Kernel Shapley Effects}\label{sec:kernanovashap}
This section discusses the available techniques for interpreting higher-order or cooperative effects that input variables exhibit on the output distribution that could be compared against the conditional indices given in Section \ref{sec:olo}.  There are two distinct cases that arise, (1) when the inputs are independent from each other and (2) when correlations/dependencies exist between the inputs.  When the inputs are independent, a kernel-ANOVA decomposition can be introduced (proven in \cite{Daveiga2021}).
\begin{theorem}[Kernel ANOVA Decomposition]\label{theorm:KANOVA} Assume the input variables are independent and $\forall R \subseteq \underline{n}$ and $\mathbf{x}_R \in \mathcal{X}_R$ $\mathbb{E}_{\mathbf{y} \sim f_{\mathbf{Y}|\mathbf{X}_R=\mathbf{x}_R}}[k(\mathbf{y},\mathbf{y})|\mathbf{X}_R] < \infty$.  Let,
	\begin{equation}
		\mathcal{S}_R^k = \sum_{U \subseteq R} (-1)^{|R|-|U|}\beta_U^k.
	\end{equation}
	and the total MMD based sensitivity can be decomposed as,
	\begin{equation}
		\sum_{R \subseteq \underline{n}} \mathcal{S}_R^k = 1 = \beta_{\underline{n}}^k.
	\end{equation}
\end{theorem}
This result can be thought of as a kernelized-Sobol index, which quantifies the influence of the input variables $R$ with all lower-order effects removed.  For instance, the first order values remain unchanged, $\mathcal{S}_i^k = \beta_i^k$ \hspace{0.1cm} $\forall i \in \{1,...,n\}$, while the second order ANOVA effect simply subtracts out the first-order indices, $\mathcal{S}_{(i,j)}^k = \beta_{(i,j)}^k - \beta_i^k - \beta_j^k, \forall i,j \in \{1,...,n\} \text{ and } i \neq j$.
This decomposition quantifies the output uncertainty caused by the inputs into separate individual and cooperative components.  Importantly, \cite{Daveiga2021} remarks that this is the first time such a decomposition has been demonstrated for a moment-independent sensitivity measure.  Note that the conditional indices can be expanded in terms of the ANOVA effects, assuming input variable independence,
\begin{equation}\label{eq:condanovarel}
	\beta_{i|A}^k = \sum_{\substack{U \subseteq A \\ S = U \cup \{i\}}}\mathcal{S}_S^k,\hspace{0.2cm} \forall A \subset \underline{n}, i \notin A.
\end{equation}
Thus, $\beta_{i|A}^k$ is the sum of all cooperative effects between $X_i$ and the input set $\mathbf{X}_A$, including the individual $\beta_i^k$ index ($U = \emptyset$).  For example, with second-order indices it is observed that $\beta_{i|j}^k=\beta_i^k+\mathcal{S}_{(i,j)}^k$.

When the assumption of input independence is no longer satisfied, a common approach is to utilize Shapley values \cite{Owen2017,Song2016,Iooss2019}.  The use of Shapley effects for GSA was first proposed in \cite{Owen2014}, borrowing a well-known solution from game theory for fairly attributing the worth created in a team effort to its individual members \cite{Shapley1953}.  In \cite{Owen2014}, Shapley effects were constructed from variance-based indices, with higher-order cooperative effects being shared equally among their underlying input variables.  The works of \cite{Owen2017,Song2016,Iooss2019} show these Shapley effects could provide meaningful insights for correlated systems.  This concept was extended in \cite{Daveiga2021} to generate a kernel-based Shapley indicator.
\begin{definition}[Kernel Shapley Effects]\label{def:kernshap} For a system with $n$ input variables, the kernel-shapley effect for variable $X_i, i \in \{1,..n\}$, is given by:
	\begin{equation}\label{eq:betashap}
		\text{Sh}_i = \frac{1}{n}\sum_{\substack{A \subset \underline{n} \\ i\notin A}} {n-1\choose|A|}^{-1}\big(\beta_{A \cup {i}}^k - \beta_{A}^k\big).
	\end{equation}
This notation assumes $\beta_\emptyset^k = 0$ for simplicity.  This definition gives the total uncertainty decomposition as,
\begin{equation}
	\sum_{i=1}^n \text{Sh}_i = 1.
\end{equation}	
\end{definition}
By applying Lemma \ref{lem:condlem} to Eq. \ref{eq:betashap}, the Shapley effects can be expressed as the sum of all conditional indices for the variable of interest,
\begin{equation}\label{eq:shapcond}
	\text{Sh}_i = \frac{1}{n}\sum_{\substack{A \subset \underline{n} \\ i\notin A}} {n-1\choose|A|}^{-1}\beta_{i|A}^k.
\end{equation}
where $\beta_{i|\emptyset}^k = \beta_i^k$.  Consequently, the Shapley effect will only be zero if all the conditional indices for learning input $i$ are zero.  Kernel Shapley values will be expanded further in Section \ref{sec:easy} with an example.

\section{Case Studies}\label{sec:casestudies}
In this section four distinct systems are used to demonstrate the previously discussed procedures.  The first two examples in Sections \ref{sec:easy} and \ref{sec:affine} are simple test cases that provide empirical comparisons between the methods discussed in Sections \ref{sec:kernelgsa} and \ref{sec:compmeth}.  After this, more practical systems are considered.  The third example in Section \ref{sec:api} is a model of a continuous flow reactor with highly correlated inputs.  The final example is a physics-based model for a lithium-ion battery.  The input kernel hyperparameters of the CME were determined through a 5-fold cross-validation procedure of \cite{Grunewalder2012b}, optimized by a Nelder-Mead simplex algorithm built into \textsc{Matlab}R2020b \cite{Lagarias1998}.

\subsection{Linear System with Variance-Based Measure}\label{sec:easy}
In this section will consider the analytical results for a linear system in order to compare the Shapley effects and OLS decomposition.  For this purpose we select the output kernel $k(\mathbf{y},\mathbf{y}') = \mathbf{y}\mathbf{y}'$, so that the corresponding $\beta^k$-indicator is equivalent to the variance-based measure\cite{Barr2022,Daveiga2021},
\begin{equation}\label{eq:varbeta}
	\beta_R^{\text{Var}}=\frac{\text{Var}\big[\mathbb{E}_{\mathbf{Y}|\mathbf{X}_R}[\mathbf{Y}|\mathbf{X}_R]\big]}{\text{Var}[\mathbf{Y}]}.
\end{equation} 
\begin{example}[Analytical Linear System]\label{example1}  Consider a system with three inputs, $\mathbf{X}_{\underline{n}} = (X_1,X_2,X_3)$ and the following input-output map,
	\begin{equation}\label{eq:ex1}
		Y = 3X_1+2.1X_2+1.9X_3.
	\end{equation}
Suppose the inputs are joint normally distributed,  $\mathbf{X}_{\underline{n}} \sim \mathcal{N}(\nu_X,\Sigma_X)$, with mean-vector and covariance matrix being
\begin{equation} \begin{split} \label{inputspread1}
		\text{ }\text{ }\nu_X &= 
		\begin{bmatrix}
			0 \\
			0 \\
			0
		\end{bmatrix} \text{, and }\text{ }\Sigma_X =
		\begin{bmatrix}
			1 & 0 & 0 \\
			0 & 1 & 0.8\\
			0 & 0.8 & 1
		\end{bmatrix},
\end{split}\end{equation}
\end{example}\pagebreak

This system is a special case of one presented in \cite{Iooss2019}, which contains formulas to calculate the exact results for the variance-based method.  Table \ref{tab:ex1s4} gives the analytical OLS decomposition and Shapley effects for Example \ref{example1}.  The first-order indices are given by the column labeled $\beta_R^{\text{Var}}$ in the OLS.  As such, note that the first-order variance-based measure would give the following order of prioritization for the variables: $X_2 > X_3 > X_1$.  In contrast, utilizing the Shapley effects for factor prioritization would instead suggest that $X_1$ is the most influential, with the order of importance being $X_1>X_2>X_3$.  

\begin{table}[h]
	\begin{center}
		\caption{OLS Decomposition and Shapley effects for Example \ref{example1}.\label{tab:ex1s4}}
		\begin{tabular}{c c c c c}
			\hline
			$X_R$ & $\beta_R^{Var}$ & $\beta_{R|2}^{Var}$ & $\beta_{R|(1,2)}^{Var}$ & Step-wise Sum\\
			\hline
			2 & 0.560 & - & - & 0.560\\
			1 & 0.384 & 0.384 & - & 0.944 \\
			3 & 0.548 & 0.056 & 0.056 & 1\\
			\hline
			\multicolumn{5}{c}{Shapley Values}\\
			\hline
			& $X_1$ & $X_2$ & $X_3$ & Sum\\
			$\text{Sh}^k_R$ & 0.384 & 0.314 & 0.302 & 1\\
			\hline
\end{tabular}\end{center}\end{table}

Thus, when correlations exist, the splitting of contributions between shared correlative/cooperative effects can lead to a discrepancy between first-order and Shapley indices in terms \textit{factor prioritization}.  The contradiction that arises here is that the $\text{Sh}^k_R$ index reports the variable that would cause the smallest reduction to output variance if individually fixed, with $X_1$ as the most influential.  This conclusion stems from the equitable principle of Shapley effects that divides the correlative influence of inputs $X_2$ and $X_3$ between the two.  The distinction involved is important, as the analysis goals should influence which GSA tools are utilized.  Shapley values have proven to be very powerful for GSA in terms of \textit{factor fixing}.  Specifically, it has been established that a zero first-order variance-based sensitivity measure is insufficient allow for factor fixing as higher-order effects may still be present \cite{Iooss2019}.  

Shapley values, even in the case of correlated inputs, can account for higher-order influences.  Thus, $Sh^k_i=0$ would indicate that an input variable can be fixed without influencing the output uncertainty \cite{Iooss2019}.  The setting of factor prioritization is more complicated.  As stated by Iooss et al. \cite{Iooss2019}, ``However, the [factor prioritization] Setting is not precisely achieved because we cannot distinguish the contributions of the main and interaction effects in a Shapley effect for the case of independent inputs."  As shown in this example, this circumstance is true for dependent inputs as well, because splitting the importance of two correlated variables can imply that a certain input factor contains less information about the output then it truly does.  To fully understand factor prioritization the ability to parse out the mutual information between two inputs when dependencies exist is necessary.  Furthermore, in general it is not the case that the introduction of correlations will strictly decrease the Shapley indicator, as the nature of the input-output map and input dependency structure both play a role \cite{Iooss2019,Plischke2021}.

With the remarks above in mind, the conditional indices can elucidate these more complex situations.  We now refer to the OLS decomposition of Table \ref{tab:ex1s4}, with the variables ordered in column $X_R$ by the OLS.  The first-order analysis reports that if only one variable could be learned (or fixed) then the order of priority is $X_2>X_3>X_1$.  However, the $\beta_{R|2}^{\text{Var}}$ column of Table \ref{tab:ex1s4} reveals that if two variables could be prioritized then $X_1$ should be prioritized ahead of $X_3$, despite $\beta_{3}^{\text{Var}}>\beta_{2}^{\text{Var}}$.  Thus, a large reduction from $\beta_{3}^{\text{Var}}$ to $\beta_{3|2}^{\text{Var}}$ is observed, which indicates that reducing the source of uncertainty from $X_2$ will naturally account for much of output variation provided by $X_3$.  This behavior arises due to the significant correlation between $X_2$ and $X_3$. In this circumstance, the OLS quantifies the mutual information that inputs $X_2$ and $X_3$ share in the output uncertainty and prioritizes $X_2$ which will account for a larger portion of uncertainty.  The OLS gives an order of factor prioritization based on learning the inputs step-by-step.  Fixing $X_2$ would cause 56.0\% reduction of output variation, additionally accounting for $X_1$ would cause a 38.4\% reduction of output variation, and if $X_3$ can be determined last it will account for the remaining 5.6\% of the outputs variance. 

\subsection{Affine Test Case with Moment-Independent Measure}\label{sec:affine}
This section provides a numerical study of an affine system to help understand the convergence behavior of the procedures provided in Sections \ref{sec:CME} and \ref{sec:compnum}.
\begin{example}[Affine System]\label{example2} Consider a system with four inputs, $\mathbf{X}_{\underline{n}} = (X_1,X_2,X_3,X_4)$, and a scalar output given by,
	\begin{equation}\label{eq:examp2}
		Y = X_1 + X_2 + 2X_3.
	\end{equation}
The inputs are joint normally distributed,  $\mathbf{X}_{\underline{n}} \sim \mathcal{N}(\nu_X,\Sigma_X)$, with mean-vector and covariance matrix being
\begin{equation} \begin{split} \label{inputspread2}
	\text{ }\text{ }\nu_X &= 
	\begin{bmatrix}
		0 \\
		0 \\
		0 \\
		0
	\end{bmatrix} \text{, and }\text{ }\Sigma_X =
	\begin{bmatrix}
		1 & 0.1 & 0 & 0 \\
		0.1 & 1 & 0.3 & 0 \\
		0 & 0.3 & 1 & 0.9 \\
		0 & 0 & 0.9 & 1
	\end{bmatrix},
\end{split}\end{equation}
respectively.  For the output kernel we follow the heuristic of \cite{Barr2022}, and utilize the Gaussian RBF kernel with the bandwidth parameter set as the standard deviation of the output, $\sigma = 2.7203$.
\end{example}
The closed-form solutions for any general affine system with joint normal input variables are given in \cite{Barr2022} (Supplemental Theorem SM5.1).  Following these formulas the exact $\beta_R^k$-index values and ISF functions can be calculated.  We report the OLS decomposition and kernel Shapley values of this system in Table \ref{tab:ex2s}.  These results follow similar logic to the systems presented in \ref{sec:easy}. Specifically $X_3$ can be seen to be the most important variable based on first-order $\beta_R^k$-indices.  The conditional effects reveal that much of the actual information provided by $X_4$ is also captured by $X_3$ as $\beta_{4|3}^{k}$ decreases greatly compared to $\beta_{4}^{k}$.  In fact, following the OLS, once inputs $(X_1,X_2,X_3)$ are learned the exact output can be determined, shown by the stepwise-sum being equal to 1.  In comparison, the kernel-Shapley effects show that no input could be fixed without changing the system response.  This includes $X_4$ despite not being in the map of Eq. \ref{eq:examp2}, due to correlations.  The indices given to calculate the OLS are in agreement with this behavior as $X_4$ has a non-zero index presented in the OLS decomposition.

\begin{table}[h]
	\begin{center}
		\caption{OLS Decomposition and Shapley effects for Example \ref{example2}.\label{tab:ex2s}}
		\begin{tabular}{c c c c c c}
			\hline
			$X_R$ & $\beta_R^{k}$ & $\beta_{R|3}^{k}$ & $\beta_{R|(1,3)}^{k}$ & $\beta_{R|(1,2,3)}^{k}$ & Step-wise Sum\\
			\hline
			3 & 0.5221 & - & - & - & 0.5221\\
			1 & 0.0812 & 0.2338 & - & - & 0.7560\\
			2 & 0.2223 & 0.2136 & 0.2440 & - & 1.0000\\
			4 & 0.2573 & 0.0656 & 0.0944 & 0 & 1.0000\\
			\hline
			\multicolumn{6}{c}{Shapley Values}\\
			\hline
			& $X_1$ & $X_2$ & $X_3$ & $X_4$ & Sum\\
			$\text{Sh}^k_R$ & 0.1860 & 0.2382 & 0.3864 & 0.1894 & 1\\
			\hline
\end{tabular}\end{center}\end{table}

In order compare the numerical methods all 15 unconditional indices were numerically calculated from single data sets with sizes $N\in \{100,250,500,750,1000\}$.  Each analysis was repeated 30 times for both the CME method and nearest-neighbor approach. For the CME, a Gaussian RBF kernel was used for each input set.  For the nearest-neighbor sub-sample method, $N_A=N$ was utilized.  This was selected based on numerical tests (not presented here) that showed a decrease in the error of the estimates as $N_A$ increased, which is in agreement with the principles presented in \cite{Broto2020}.  

\begin{figure}[t] 
	\center
	\includegraphics[width=\linewidth]{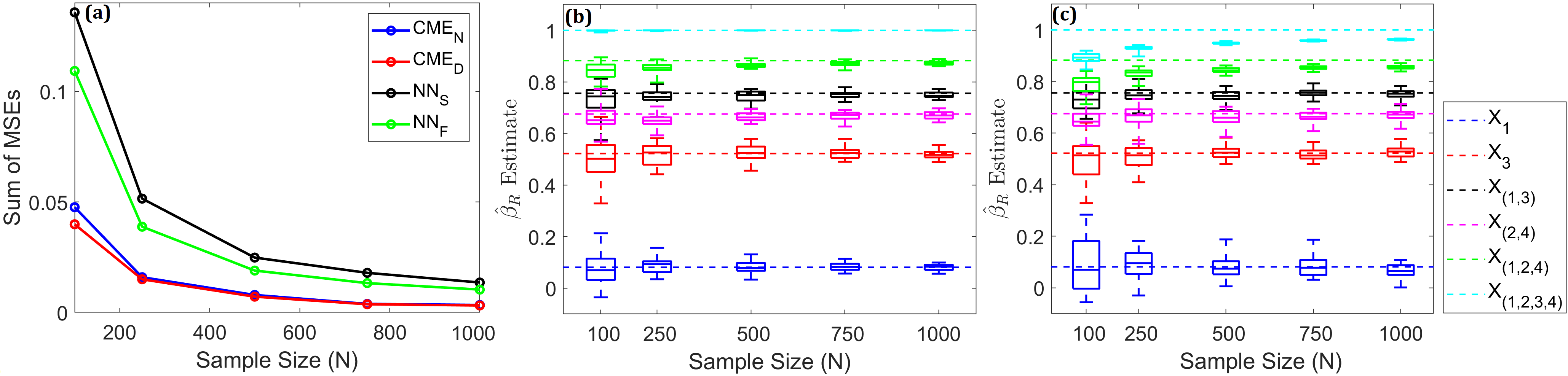}
	\caption{(a) The sum of the 15 MSEs calculated for the four estimation procedures.  The labels CME and NN are for the conditional embedding and nearest neighbor approaches, respectively. CME with subscripts N and D are for the norm-based and distance-based estimates (Eq. \ref{eq:Fest}), respectively.  NN with subscripts S and F are for the sub-sample and full-sample based estimates (Eq. \ref{eq:NNjoint}), respectively.   (b) Box plots for the 30 replicates of the norm-based CME estimate ($\hat{\beta}_R^{k,N}$, Eq. \ref{eq:Fest}) with respect to the total data sample size.  The dashed lines represent analytical values that correspond the box plots of the same color.  (c) A box plot for full-sample nearest neighbor estimator ($\hat{\beta}^{k,F}_R$, Eq. \ref{eq:NNjoint}) following the same format as sub-figure (b).}
	\label{fig:AffineErrors}
\end{figure}

The results are reported in Figure \ref{fig:AffineErrors}.  For each of the 15 $\beta_R^k$ indices the mean squared error (MSE) was calculated, averaging across the 30 repeated analyses for each sample size.  For brevity, the MSEs for all 15 $\beta_R^k$ values were summed and plotted in Figure \ref{fig:AffineErrors}(a).  The Figures \ref{fig:AffineErrors}(b) and \ref{fig:AffineErrors}(c) give a more detailed analysis for some selected $\beta_R^k$ indices by presenting box plots of the numerical methods across the 30 repetitions.  In terms of total errors, Figure \ref{fig:AffineErrors}(a), reveals that the CME outperforms the nearest neighbor approach in this example.  Interestingly, the nearest-neighbor with the full input-output ($\hat{\beta}^{k,F}_R$, Eq. \ref{eq:NNjoint}) sample has a slight improvement compared to the bootstrapped sub-sample procedure $(\hat{\beta}^{k,S}_R$, Eq. \ref{eq:NNjoint}).

Figure \ref{fig:AffineErrors}(b) shows box plots for only the norm-based CME result $(\hat{\beta}^{k,N}_R$, Eq. \ref{eq:Fest}) since the results are very similar to the distance-based $(\hat{\beta}^{k,D}_R$, Eq. \ref{eq:Fest}) approach for $N \geq 250$. Figure \ref{fig:AffineErrors}(c) gives the full-sample nearest-neighbor estimate ($\hat{\beta}^{k,F}_R$, Eq. \ref{eq:NNjoint}) because it outperformed the sub-sample method ($\hat{\beta}^{k,S}_R$, Eq. \ref{eq:NNjoint}).  The consistency of the CME routine is observed as the estimates converge towards the analytical values for each input set as $N$ increases. The same behavior is seen for the first and second order indices calculated by the nearest-neighbor technique.  However the third and fourth order indices do have bias present, as predicted by \cite{Deb2020}, as the numerical estimate is lower than that of the analytical result.  While not shown, the same behavior is found with the sub-sample estimator.  This outcome is in-line with \cite{Daveiga2021}, which found the nearest neighbor sub-sampling methodology to still contain unresolved bias, like the full-sample estimate.

\begin{figure}[t] 
	\center
	\includegraphics[width=\linewidth]{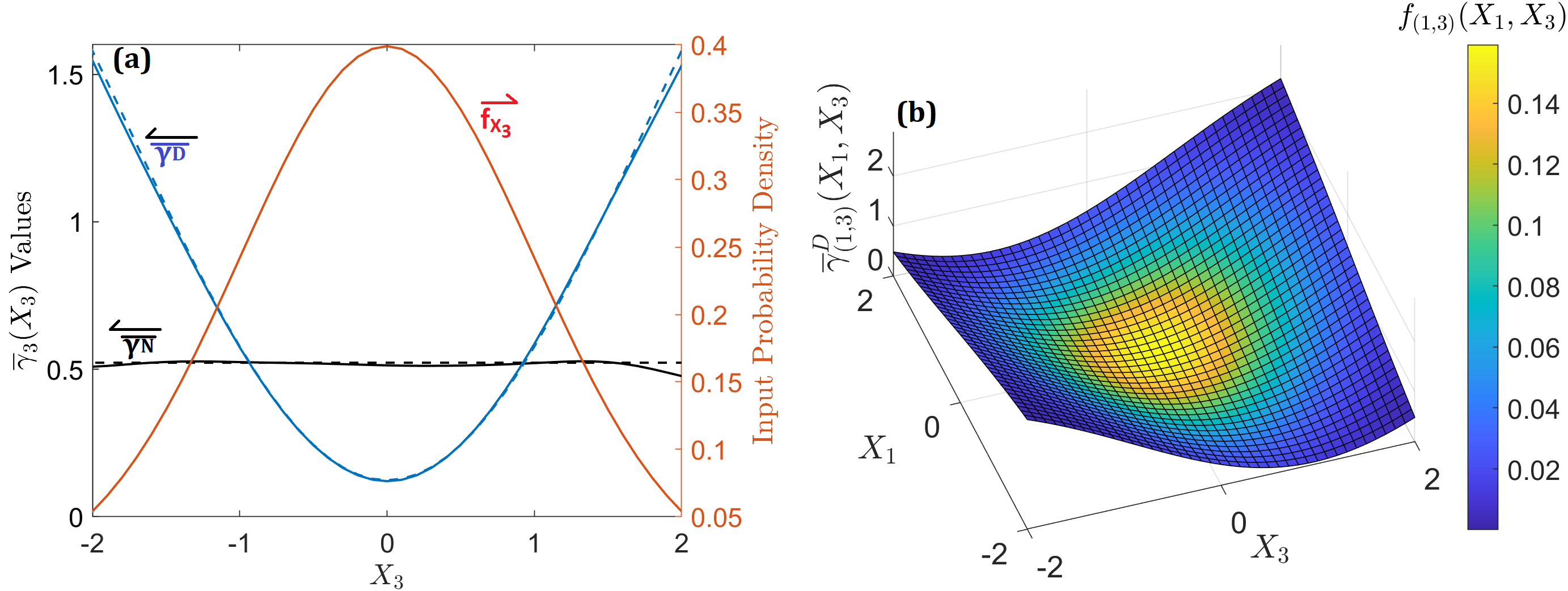}
	\caption{(a) A plot with the $\overline{\gamma}^N$ in black and  $\overline{\gamma}^D$ function in blue for input $X_3$.  The true analytical functions are given by dashed lines with the numerical estimates given by the solid lines.  The red plot is the probability distribution for input $X_3$. The arrows above the labels indicate which axis range is used for the corresponding function. (b)  A plot of the analytical $\overline{\gamma}^D$ function for inputs $X_1$ and $X_3$, the color-plot is used to represent the underlying joint probability distribution of the corresponding inputs.}
	\label{fig:AffineISF}
\end{figure}

Lastly, two select ISFs are given in Figure \ref{fig:AffineISF}.  Figure \ref{fig:AffineISF}(a) reports the ISFs for $X_3$. The input distribution of $X_3$, labeled $f_{X_3}$, is given to aid in the interpretation by providing the prior assigned probabilities of $X_3$.  For both ISFs, the average numerical function determined from the 30 repetitions of $N=1000$ data points is given by the solid line.  The numerical and analytical results are in good agreement, with the error increasing as the functions move to values with lower probabilities.  The function $\overline{\gamma}_3^N$ with the Gaussian RBF effectively measures the over-all spread of the output data, with larger function values representing less variance in the output data.   As described in \cite{Barr2022}, the unconditional and conditional output distributions will both be normally distributed.  Remarkably, the input-output relationship is homoscedastic, meaning the conditional output variance is constant across the input range, which is reflected by $\overline{\gamma}_3^N$ being a constant line.

The function $\overline{\gamma}_3^D$ gives a measure of the difference between conditional and unconditional output distribution.  Noting that both output distributions are normally distributed with constant variance, then the difference between the two can be determined just from the difference between their means. The conditional output mean, unlike the conditional variance, does depend on the value of the fixed input variables. This behavior is seen with $\overline{\gamma}_3^D$ as the minimum is observed at the mean, $X_3=0$, and increases when $X_3$ deviates from the average.  In general these functions reference entire output conditional and unconditional distributions, not just the first two statistical moments, but the fact that both distributions are joint normal with fixed variance make the interpretation appropriate in such terms.  

Figure \ref{fig:AffineISF}(b) gives the true $\overline{\gamma}_R^D$ function for input sets $X_R = (X_1,X_3)$ to provide insights into second-order interactive effects.  The underlying joint probability distribution is given by the color-plot, so inferences based on the probabilities of events may also be made.  These account for both the direct input-output mapping given by Eq. \ref{eq:examp2} and the correlations between inputs.  Noting that in Example \ref{example2} $\mathbb{E}_{Y\sim f_{Y}}[Y] = 0$, then through similar logic as before, the largest discrepancies are observed for joint input values that give a conditional output mean furthest from 0. With Figure \ref{fig:AffineISF}(b) this is observed when $X_1$ and $X_3$ are large in magnitude, but agree in sign.  This behavior may be understood as the $X_1+2X_2$ terms would combine to give a conditional output mean different from zero.  When the two inputs are opposite in sign the $X_1+2X_2$ terms could cancel out to still give a conditional mean near zero, which is why lower values are observed when the two are opposite in sign but still large in magnitude. 

To conclude, there are three main results this example demonstrates.  First, it is an empirical demonstration of the improvement in numerical estimation that the CME procedure exhibits in terms of convergence rates and bias.  Additionally, the conditional indices can provide insights into cooperative input variable effects, even in the presence of correlations, with the OLS being a potential approach to factor prioritization.  Lastly, the ISFs determined from a single data set are capable of providing insights into how inputs marginally influences the output distribution.

\subsection{Continuous Flow Chemical Reactor}\label{sec:api}
For the next input-output system we consider the model of a continuous-flow reactor for the synthesis of aminopyrimidine presented in \cite{Reizman2012} as a practical system with highly correlated inputs.   The goal of \cite{Reizman2012} was to couple an online analysis method with an experimental automated continuous flow system to address the challenge of modeling and optimizing product yield or selectivity in multi-step reaction networks with little \textit{a priori} reaction information.   For additional insights, a first-order moment-independent analysis was done in the work of \cite{Xie2019}.  

The model of \cite{Reizman2012} assumes all the reaction kinetics are second-order, occurring in an ideal plug flow reactor.  The mechanism is given by the following chemical reactions:
\begin{equation}\begin{split}\label{eq:chemeq}
	 &\ce{A +B ->[k1]C}\\
	 &\ce{A +B ->[k2]D}\\
	 &\ce{C +B ->[k3]E}\\
	 &\ce{D +B ->[k4]E}
\end{split}\end{equation}
where the rate constants $k_i, i \in \{1,...,4\}$ are of the form,
\begin{equation}\label{eq:kp}
	k_i = A_i\text{exp}\Bigg[-\frac{E_{A_i}}{RT} \Bigg].
\end{equation}
The variables $A_i$ and $E_{A_i}$ represent the pre-exponential factors and activation energies, respectively, which are the uncertain inputs.  Further, $T$ is a constant temperature as the reactor is assumed to be isothermal and $R$ is the ideal gas constant.  The reactions of Eq. \ref{eq:chemeq} describe the nucleophilic aromatic substitution reactions of 2,4-dichloropyrimidine (A) and morpholine (B) in ethanol to create both the target product of 2-substituted aminopyrimidine (D) and a byproduct of 4-substituted aminopyrimidine (C).  Both C and D can react again with B to create a 2,4-substituted side-product (E).  For a more complete description of the reactions, experimental determination of parameters and discussion on the pharmaceutical significance of the aminopyrimidine products, see \cite{Reizman2012}.  

The chemical reactions of Eq. \ref{eq:chemeq} can be expressed as a system of ordinary differential equations,
\begin{equation}\begin{split}\label{eq:CFRODE}
\frac{\text{d}[A]}{\text{d}t} &= -k_1[A][B] - k_2[A][B]\\
\frac{\text{d}[B]}{\text{d}t} &= -k_1[A][B] - k_2[A][B] - k_3[B][C] - k_4[B][D]\\
\frac{\text{d}[C]}{\text{d}t} &= k_1[A][B] - k_3[B][C]\\
\frac{\text{d}[D]}{\text{d}t} &= k_2[A][B] - k_4[B][D]\\
\frac{\text{d}[E]}{\text{d}t} &= k_3[B][C] + k_4[B][D]
\end{split}\end{equation}
where the brackets are used to represent the concentration of a given species.  The reaction time runs from $t=0$ to $t=t_{res}$, where $t_{res}$ is the reaction residence time in the continuous flow reactor.  The model of \cite{Reizman2012} was developed considering different temperatures, residence times and initial reactant concentrations.  Following \cite{Xie2019}, we analyzed a single case that was explored in \cite{Reizman2012} so that T, $t_{res}$, and initial concentrations of A and B (labeled $[A]_0$ and $[B]_0$ respectively) were fixed to the values given in Table \ref{tab:CFRparam}.  The uncertain input variables were the 8 parameters defined by Eq. \ref{eq:kp} and are labeled $X_{i}$ in Table \ref{tab:CFRparam}.  Following \cite{Xie2019}, the inputs were assumed to have normal marginal distributions.  The means and standard deviation of each input are given in Table \ref{tab:CFRparam} as well.
\begin{table}[h]\begin{center}
		\caption{Parameters and input distribution uncertainties for the continuous flow reactor model.}\label{tab:CFRparam}
		\begin{tabular}{c c c c}
			\hline
			Parameter (Units) & Input Label & Mean/Nominal Value & Standard Deviation\\
			\hline
			$[A]_0$ (M) & - & 0.150 & -\\
			$[B]_0$ (M) & - & 0.375 & -\\
			$[C]_0$ (M) & - & 0 & -\\
			$[D]_0$ (M) & - & 0 & -\\
			$[E]_0$ (M) & - & 0 & -\\
			$T$ (K) & - & 373.15 & -\\
			$t_{res}$ (s) & - & 1200 & -\\
			$R$ ($\text{kJ}\text{mol}^{-1}\text{K}^{-1}$) & - & 0.008314 & -\\
			$\text{log}_{10}(A_1)$ ($\text{M}^{-1}\text{s}^{-1}$) & $X_1$ & 3.4 & 0.1 \\
			$E_{A_1}$ (kJ/mol) & $X_2$ & 27.0 & 0.6\\
			$\text{log}_{10}(A_2)$ ($\text{M}^{-1}\text{s}^{-1}$) & $X_3$ & 3.5  & 0.1 \\
			$E_{A_2}$ (kJ/mol) & $X_4$ & 32.1 & 0.6\\
			$\text{log}_{10}(A_3)$ ($\text{M}^{-1}\text{s}^{-1}$) & $X_5$ & 4.9 & 0.2 \\
			$E_{A_3}$ (kJ/mol) & $X_6$ & 60.0  & 1.6 \\
			$\text{log}_{10}(A_4)$ ($\text{M}^{-1}\text{s}^{-1}$) & $X_7$ & 3.0 & 0.2 \\
			$E_{A_4}$ (kJ/mol) & $X_8$ & 45.0  & 1.7 \\
			\hline
\end{tabular}\end{center}\end{table}

The output of interest is the final concentration of $D$ produced, $[D]$ at $t_{res}$.  Following \cite{Xie2019}, two analyses were performed considering two input distributions.  The first was utilizing the normal distributions of Table \ref{tab:CFRparam} assuming independence between the inputs.  The second input sampling method considered a dependence structure of the inputs, with the same marginal distributions given in Table \ref{tab:CFRparam}.  The correlation matrix of the inputs, determined from data in \cite{Reizman2012}, is provided in Table \ref{tab:corrmatrix}, which reveals strong input-input dependencies.   Based on the reasoning of \cite{Xie2019}, a Gaussian copula with Table \ref{tab:corrmatrix} as the underlying correlation matrix was assumed. In total, two first-order moment independent sensitivity analyses were performed, with the first ignoring correlations by assuming input independence and a second which modeled the input correlations with a Gaussian copula.  Performing both analyses provides insights into how the correlations influence the resulting uncertainty and sensitivity analysis and additionally allows for a complete comparison of results with \cite{Xie2019}. A full-order sensitivity analysis was additionally performed on only the system with input correlations, presented further below.

\begin{table}[h]\begin{center}
		\caption{Correlation matrix for input variables in continuous flow reactor, from data of \cite{Reizman2012}}\label{tab:corrmatrix}
		\begin{tabular}{c| c c c c c c c c}
			\hline
			 & $X_1$ & $X_2$ & $X_3$ & $X_4$ & $X_5$ & $X_6$ & $X_7$ & $X_8$\\
			\hline
			$X_1$ & 1.000 & 0.997 & 0.976 & 0.968 & -0.002 & -0.003 & 0.000 & 0.000\\
			$X_2$ & 0.997 & 1.000 & 0.976 & 0.973 & -0.003 & -0.003 & 0.000 & 0.000\\
			$X_3$ & 0.976 & 0.976 & 1.000 & 0.997 & -0.006 & -0.006 & 0.000 & 0.000\\
			$X_4$ & 0.968 & 0.973 & 0.997 & 1.000 & -0.007 & -0.007 & 0.000 & 0.000\\
			$X_5$ & -0.002 & 0.003 & -0.006 & -0.007 & 1.000 & 1.000 & -0.008 & -0.008\\
			$X_6$ & -0.003 & 0.003 & -0.006 & -0.007 & 1.000 & 1.000 & -0.008 & -0.008\\
			$X_7$ & 0.000 & 0.000 & 0.000 & 0.000 & -0.008 & -0.008 & 1.000 & 1.000\\
			$X_8$ & 0.000 & 0.000 & 0.000 & 0.000 & -0.008 & -0.008 & 1.000 & 1.000\\
			\hline
\end{tabular}\end{center}\end{table}

\begin{figure}[t] 
	\center
	\includegraphics[width=\linewidth]{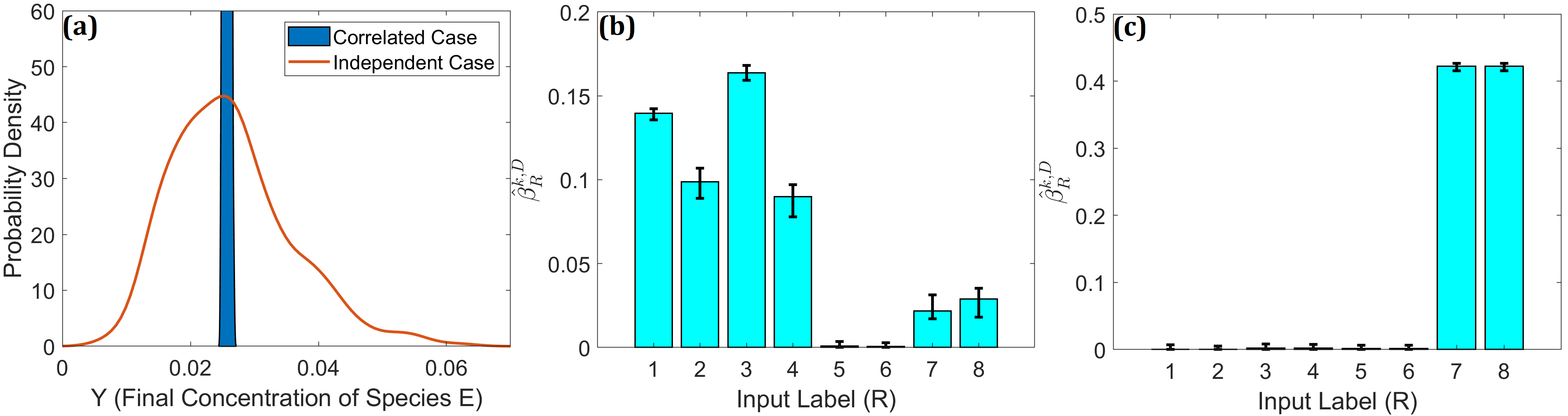}
	\caption{(a) Output probability distributions generated by a kernel density estimation of the data.  (b) Bar plots for first-order $\beta_R^k$ indices assuming an independent input distribution and (c) utilizing a Gaussian copula dependency structure.  Error bars give the maximum and minimum estimates found across the 30 repetitions.}
	\label{fig:APIuncertainty}
\end{figure}

For both cases, $N=1000$ input-output samples were generated and a first-order sensitivity analysis was performed with the CME estimators.  The output kernel was the Gaussian RBF with bandwidth set as the output standard deviation. The input kernel was also set to be the Gaussian RBF where the bandwidth was determined via cross-validation.  The cross-validation was done over 30 replicates, being performed on randomly shuffled subsets of 800 data points of the given original samples.  Results for both first-order uncertainty and sensitivity analysis are reported in Figure \ref{fig:APIuncertainty}.  Figure \ref{fig:APIuncertainty}(a) shows the output probability distributions for both data sets, indicating that the correlations have a significant role upon the resulting uncertainty in the systems output.  When the input-input dependencies are considered the total uncertainty in the output is greatly reduced ($\sim$ 30 fold reduction in output variance). Further, the corresponding sensitivity analysis changes dramatically at first-order.  Figure \ref{fig:APIuncertainty}(b) gives the results assuming independence.  The inputs $\{X_1,X_2,X_3,X_4\}$ are found to be the most significant, while inputs $\{X_7,X_8\}$ have a weaker influence, and inputs $\{X_5,X_6\}$ are of little consequence at first-order.  In Figure \ref{fig:APIuncertainty}(c) the results with the linear correlations are shown, which are dramatically different.  Inputs $X_7$ and $X_8$ are now the most important variables, with the remaining inputs are found to be unimportant at first-order.  Both sets of results are in strong agreement with \cite{Xie2019}, which did a first-order moment-independent analysis with the method of \cite{Borgonovo2007} on the same systems.  

An interesting aspect of the correlated results of Figure \ref{fig:APIuncertainty}(c), is that much of the information that inputs $X_7$ and $X_8$ contain about the output should be shared due to their deterministic relationship.  Thus, the remaining 60\% of output uncertainty must come from cooperative effects between other inputs that are individually unimportant.  Specifically, despite the output being independent or weakly dependent (due to using a moment-independent measure) on inputs $\{X_1,...,X_6\}$, it must be conditionally dependent on these inputs in order to account for all of the remaining output variation.  To parse out this information, a full-order sensitivity analysis was done with the data that considered input correlations and the results are reported in Table \ref{tab:APIOLO}, which show the OLS decomposition and Shapley effects.
\pagebreak

For the higher-order analysis of the correlated system, a Mahalanobis kernel was selected for the input kernels due to the presence of extreme input-input dependencies, i.e. a deterministic relationship for correlation coefficients of 1.000,
\begin{equation}\label{eq:mahakern}
	k_{\mathbf{M}_R}(\mathbf{x}_R,\mathbf{x}_R')=\text{exp}\big(-\frac{(\mathbf{x}_R-\mathbf{x}_R')^T\mathbf{M}_R^{+}(\mathbf{x}_R-\mathbf{x}_R')}{2\lambda^2}\big); \sigma >0.
\end{equation} 
Here $\lambda$ is a hyperparameter determined by cross-validation and the matrix $\mathbf{M}_R$ is symmetric positive semi-definite matrix.  For the analysis, $\mathbf{M}_R$ was set to be the covariance matrix of the input subset $\mathbf{X}_R$, calculated from the randomly generated input data. The $+$ superscript represents the Moore-Penrose pseudo-inverse in order for $\mathbf{M}$ to be well-defined if it is singular (i.e., $\mathbf{X}_R = \{X_7,X_8\}$).   Note that when $|R|=1$ that $\mathbf{M}_R$ is a scalar, reducing the Mahalanobis kernel to the Gaussian RBF.  Consequently, the numerical scheme to utilize a Mahalanobis kernel for the inputs is the same as utilizing the Gaussian RBF for the first-order results and produced results in agreement to those obtained above for the same data set.

\begin{table}[t]
	\begin{center}
		\caption{OLS Decomposition and Shapley effects for the continuous flow reactor accounting for input correlations.  (0) Represents a value of zero up to the reported digits. \label{tab:APIOLO}}
		\begin{tabular}{c c c c c c c c c c}
			\hline
			$X_R$ & $\beta_R^{k}$ & $\beta_{R|7}^{k}$ & $\beta_{R|(3,7)}^{k}$ & $\beta_{R|(1,3,7)}^{k}$ &
			$\beta_{R|(1,3,4,7)}^{k}$ & $\beta_{R|(1,2,3,4,7)}^{k}$ & \multicolumn{3}{c}{Step-wise Sum}\\
			\hline
			7 & 0.417 & - & - & - & - & - & \multicolumn{3}{c}{0.416}\\
			3 & 0.003 & 0.004 & - & - & - & - & \multicolumn{3}{c}{0.421}\\
			1 & 0.001 & (0) & 0.156 & - & - & - & \multicolumn{3}{c}{0.577}\\
			4 & 0.003 & 0.003 & 0.003 & 0.023 & - & - & \multicolumn{3}{c}{0.600}\\
			2 & 0.001 & (0) & 0.100 & 0.022 & 0.401 & - & \multicolumn{3}{c}{1.000}\\
			5 & 0.003 & (0) & (0) & (0) & 0.001 & (0) & \multicolumn{3}{c}{1.000}\\
			6 & 0.003 & (0) & (0) & (0) & 0.001 & (0) & \multicolumn{3}{c}{1.000}\\
			8 & 0.417 & (0) & (0) & (0) & (0) & (0) & \multicolumn{3}{c}{1.000}\\
			\hline
			\multicolumn{10}{c}{Shapley Values}\\
			\hline
			& $X_1$ & $X_2$ & $X_3$ & $X_4$ & $X_5$ & $X_6$ & $X_7$ & $X_8$ &Sum\\
			$\text{Sh}^k_R$ & 0.157 & 0.123 & 0.149 & 0.120 & 0.002 & 0.002 & 0.224 & 0.224 & 1.000\\
			\hline
\end{tabular}\end{center}\end{table}

\begin{figure}[t] 
	\center
	\includegraphics[width=0.9\linewidth]{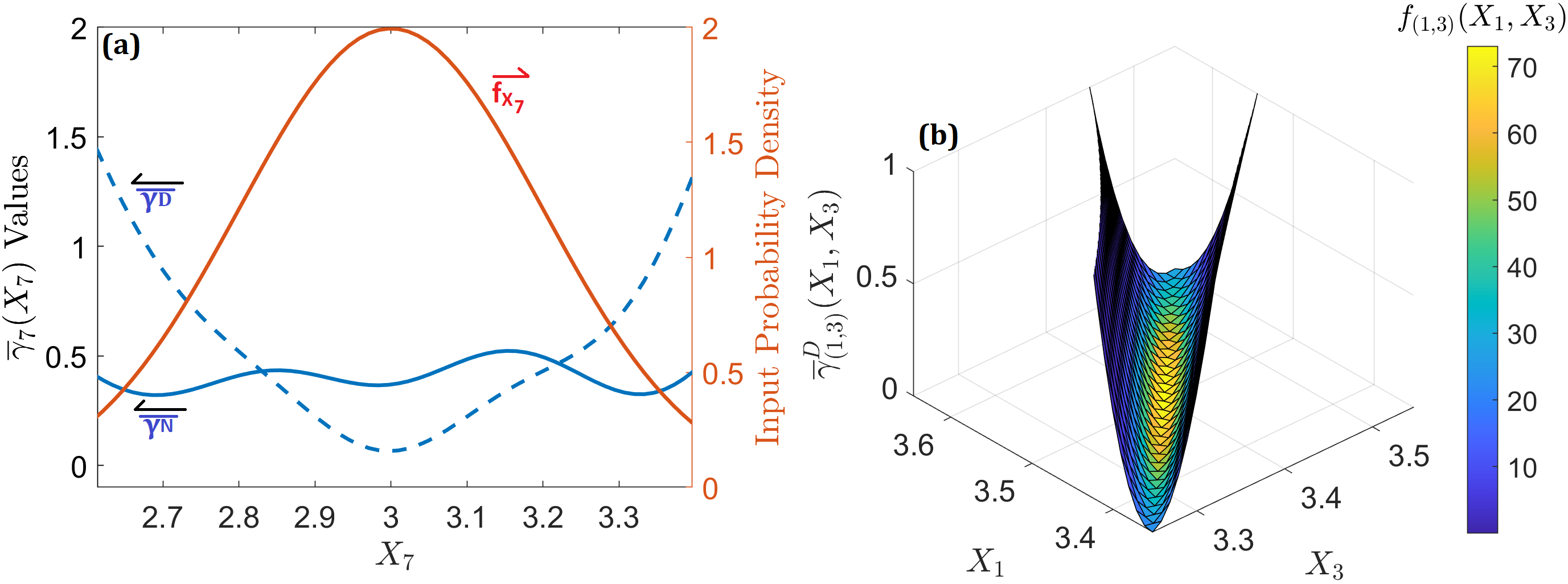}
	\caption{(a) ISFs for $X_7$ with the marginal input probability density function given in red, $f_7$.  The numeric $\overline{\gamma}^D_{X_7}$ and $\overline{\gamma}^N_{X_7}$ are given by the dashed and solid blue lines, respectively. (b) $\hat{\gamma}^D_{X_1,X_3}$ for inputs $(X_1,X_3)$.  The color-plot is used to represent the underlying joint probability distribution of the corresponding inputs.}
	\label{fig:APIISFs}
\end{figure}

As hypothesized, the conditional indices show that $X_7$ and $X_8$ hold the same information about the output variable as $\beta_7^k =\beta_8^k = 0.417$ but $\beta_{8|7}^k = 0$. Further, despite the output being weakly dependent on the inputs $\{X_1,...,X_4\}$ individually (reported in \cite{Xie2019} as well), it is not conditionally independent of these variables as $\beta_{1|(3,7)} =0.156$ and $\beta_{2|(1,3,4,7)} =0.401$.  To bolster this point, two additional indices calculated were $\beta_{(1,3)} = 0.079$ and $\beta_{(1,2,3,4)}=0.224$, which reveal significant cooperative effects for these variables. In terms of factor fixing, the OLS decomposition and Shapley effects show that inputs $X_5$ and $X_6$ can be fixed as their conditional indices and Shapley values are near zero.  After sharing the cooperative and correlative influences, the Shapley effects report inputs $X_7$ and $X_8$ still as the most important; additionally, inputs $\{X_1,X_2,X_3,X_4\}$ are of similar importance.  Note that due to the deterministic relationship either $X_7$ or $X_8$ can be used as the most influential variable, and the results are identical with $X_7$ and $X_8$ swapped, thus only the set with $X_7$ listed first is shown in Table \ref{tab:APIOLO}.  Select ISFs are plotted in Figure \ref{fig:APIISFs} to help understand the relationship between the inputs and the output distribution.

Figure \ref{fig:APIISFs}(a) gives the ISFs for the most influential variable, $X_7$.  The conditional distribution is most similar to the unconditional one near the input mean $X_7 = 3.0$ and deviates more as the input gets further from its mean, which is shown by $\overline{\gamma}_7^D$.  The ISF $\overline{\gamma}_7^N$ reports a somewhat constant decrease in data spread, with a slight oscillatory behavior.  Figure \ref{fig:APIISFs}(b) gives $\overline{\gamma}_{(1,3)}^D$ to reveal the conditional dependence effects that are prevalent.  The color-plot shows the strong dependency between these inputs, indicating that the conditional output distribution has a minimal change from the unconditional distribution along the line of correlation with differences increasing as the inputs move off the diagonal.\pagebreak

These results for the scenario with correlated inputs can be qualitatively understood by the chemical model of Eq. \ref{eq:chemeq}, and are summarized in terms of factor prioritization here:
\begin{itemize}
	\item The most influential input variables are $X_7$ and $X_8$.  These inputs are highly correlated and efforts to reduce the source of uncertainty from one will naturally reduce the source from the other.  As seen from the fourth reaction of Eq. \ref{eq:chemeq}, these parameters correspond to the consumption of the species of interest, $[D]$ at $t_{res}$, to generate byproduct E and together account for $\sim 41\%$  the output uncertainty.
	\item The next influential set of inputs are  $\{X_1,X_2,X_3,X_4\}$, which together account for $\sim22\%$ of the uncertainty.  These inputs are grouped, as despite being highly correlated, must be determined simultaneously for significant reductions in uncertainty.  These parameters give the rates of the first two reactions in Eq. \ref{eq:chemeq}.  One possible explanation for these conditional effects could be the competitive nature of the first two reactions of Eq. \ref{eq:chemeq}.  Specifically, these reactions run to completion as $A$ is consumed to produce either $C$ or $D$, so $[D]$ is directly proportional to $k_2$ and inversely proportional to $k_1$. 	Thus, if the information about these rate constants is highly correlated then only determining one variable may not provide information about the differences between rate constants.  That is, if $X_1$ is found to be larger in value then expected, then both $k_1$ and $k_2$ will increase by similar factors resulting in little change in the difference between $k_1$ and $k_2$.	
	\item The remaining $37\%$ comes between cooperative effects between the sets $\{X_1,X_2,X_3,X_4\}$ and $\{X_7,X_8\}$.
	\item The two variables $X_5$ and $X_6$ are unimportant and can be fixed.  These parameters give the rate of the third reaction of Eq. \ref{eq:chemeq}.  Assuming $B$ is in abundance then these only correspond to the the consumption of $C$ and would have no effect on $D$.  These parameters may contain uncertainty about other outputs, such as concentrations of $C$ or $E$, but they can be fixed without influencing the uncertainty of $D$.
\end{itemize}
To conclude, the ISFs and conditional indices help reveal the cooperative and correlative effects of the input variables of this model.

\subsection{Lithium-Ion Battery Model}\label{sec:battery}
In this section an analysis is performed on the partial differential equation model of a lithium-ion battery (LIB) described by Hadigol et al. \cite{Hadigol2015}.  In \cite{Hadigol2015}, an uncertainty analysis was performed, which provides the selection and justification for input distributions of 19 different uncertain inputs, assuming independence between the input variables.  In total, 3600 CPU hours were needed to generate the data, which it was made publicly available in \cite{Constantine2017}.  The 19 uncertain inputs and their distributions are given in Table \ref{tab:batteryparam}.  A key factor for the battery system that was not treated as an uncertain parameter is the discharge rate of the battery.  In \cite{Hadigol2015}, three different data sets of size $N=3600$ were generated by Monte Carlo simulation, where each data set assumed a constant discharge rate at one of three values: 0.25C, 1C, and 4C.  The term C denotes the so-called C-rate  measuring  the  rate  at  which  a  battery  discharges  from  its  full  capacity.  The authors of \cite{Hadigol2015} note that battery loading should vary across the discharge process and that treating the discharge rate as a constant for each data set is an idealized scenario.
   
\begin{table}[h]\begin{center}
		\caption{Descriptions and distributions for the uncertain parameters of the LIB model. $U(a,b)$ represents a uniform distribution with minimum $a$ and maximum $b$, while $\mathcal{N}(\nu_1,\nu_2)$ describes a normal distribution with mean $\nu_1$ and standard deviation $\nu_2$.}\label{tab:batteryparam}
		\begin{tabular}{c c c c c}
			\hline
			Parameter Label & Input Label & Units & Description & Distribution\\
			\hline
			$\epsilon_a$ & $X_1$ & - & Anode porosity & U(0.46,0.51)\\
			$\epsilon_s$ & $X_2$ & - & Separator porosity & U(0.63,0.81)\\
			$\epsilon_c$ & $X_3$ & - & Cathode porosity & U(0.36,0.41)\\
			$\text{brugg}_a$ & $X_4$ & - & Anode Bruggeman coefficient & U(3.8,4.2)\\
			$\text{brugg}_s$ & $X_5$ & - & Separator Bruggeman coefficient & U(3.2,4.8)\\
			$\text{brugg}_c$ & $X_6$ & - & Cathode Bruggeman coefficient& U(3.8,4.2)\\
			$t_0^+$ & $X_7$ & - & $\text{Li}^+$ transference number & U(0.345,0.381)\\
			$D$ & $X_8$ & $m^2\,s^{-1}$ & Salt diffusion coefficient in liquid  & U(6.75,8.25)$\times 10^{-10}$\\
			$D_a$ & $X_9$ & $m^2\,s^{-1}$ & Anode solid diffusion coefficient  & U(3.51,4.29)$\times 10^{-14}$\\
			$D_c$ & $X_{10}$ & $m^2\,s^{-1}$ & Cathode solid diffusion coefficient & U(0.90,1.10)$\times 10^{-14}$\\
			$\sigma_a$ & $X_{11}$ & $S\,m^{-1}$ & Anode conductivity & U(90,110)\\
			$\sigma_c$ & $X_{12}$ & $S\,m^{-1}$ & Cathode conductivity & U(90,110)\\
			$k_a$ & $X_{13}$ & $m^{4}\,\text{mol}\,s$ & Anode reaction rate & U(4.52,5.53)$\times 10^{-11}$\\
			$k_c$ & $X_{14}$ & $m^{4}\,\text{mol}\,s$  & Cathode reaction rate & U(2.10,2.56)$\times 10^{-11}$\\
			$r_a$ & $X_{15}$ & $\mu m$ & Anode particle size & $\mathcal{N}$(2,0.1354)\\
			$r_c$ & $X_{16}$ & $\mu m$  & Cathode particle size & $\mathcal{N}$(2,0.3896)\\
			$L_a$ & $X_{17}$ & $\mu m$ & Anode length & U(77,83)\\
			$L_s$ & $X_{18}$ & $\mu m$ & Separator length & U(22,88)\\
			$L_c$ & $X_{19}$ & $\mu m$ & Cathode length & U(85,91)\\
			\hline
\end{tabular}\end{center}\end{table}\pagebreak

For a full description of the partial differential equation model and the implications of GSA for quality control procedures of LIBs, see \cite{Hadigol2015}.  The LIB model simulates cell capacity as a function of voltage while the battery discharges, cell voltage as a function of time, and the lithium concentration in various parts of the battery with respect to time, which are all potential quantities of interest.  All simulations were run until the voltage reached a cut-off value of 2.8 V. The time to reach this threshold varied with the randomized input values.  Due to this behavior a scaled time-coordinate, $t^{*}$, was used \cite{Hadigol2015}. This was set to $t^{*}=100$ at physical time $t=0$, and the time $t^{*}=0$ was when the voltage reached 2.8 V.  In this fashion, $t^{*}$ represent a charge meter from $100\%$ to $0\%$.

\begin{figure}[t] 
	\center
	\includegraphics[width=\linewidth]{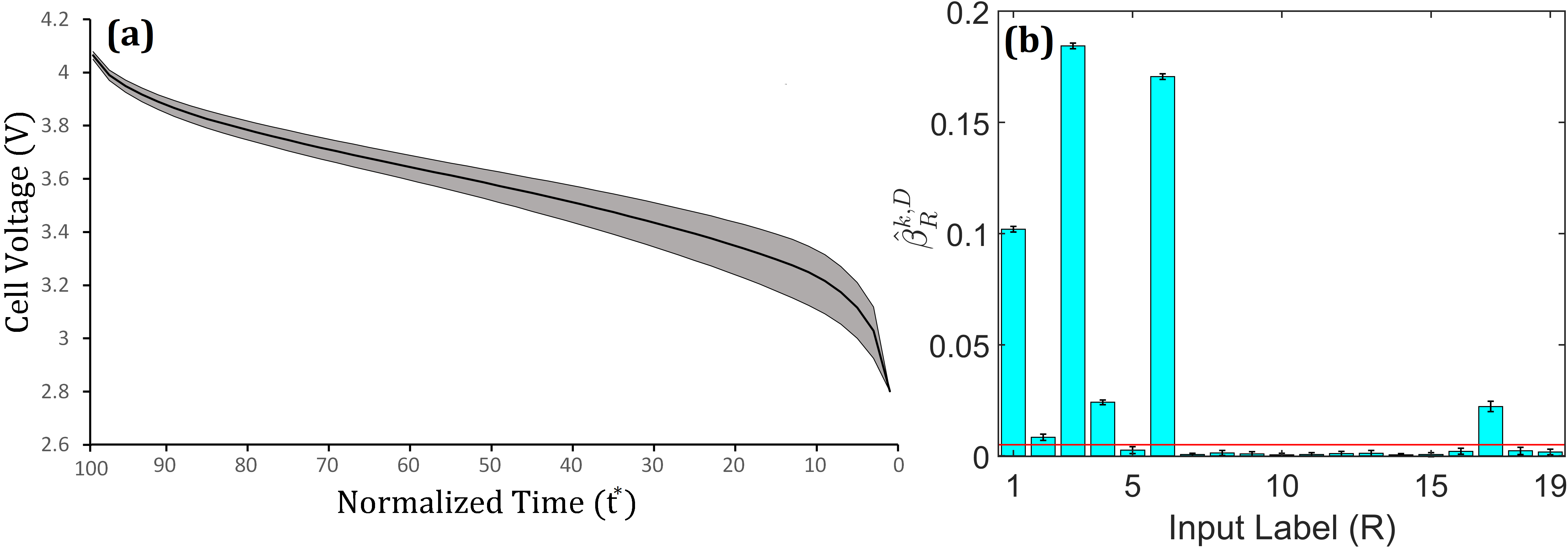}
	\caption{(a) Discharge curve of the LIB with $I=1C$ rate of discharge. The solid line represents the mean, while the shaded area gives uncertainty bounds between the 5th and 95th data quantiles.  (b) Numerical first-order sensitivity indices for the LIB, with error bars of one standard deviation across 30 shuffled cross-validation replicates.  The red-line is the $\hat{\beta}_R=0.005$ screening threshold.}
	\label{fig:BUQ1}
\end{figure}

For this analysis, we limit our focus on the data set with a discharge rate of 1C and the output to be cell voltage as a function of normalized time.  The voltage is recorded at 50 times points from the initial voltage to the final voltage of 2.8 V, making the output a 50-dimensional real vector, $\mathcal{Y} \subseteq \mathbb{R}^{50}$.  For the output kernel a Gaussian RBF was utilized where the bandwidth was set to be the square root of the trace of the covariance matrix of the multivariate outputs.  This can be calculated directly from the provided data and is based on the heuristic of \cite{Barr2022}.   For the input kernel, the data provided by \cite{Hadigol2015,Constantine2017} have the inputs normalized to the hypercube $[-1,1]^{-19}$ so that a simple choice of the Gaussian RBF is sufficient even when considering inputs with different units and scales.

\begin{table}[h]
	\begin{center}
		\caption{OLS Decomposition, Shapley effects and select kernel-ANOVA indices for the LIB model. \label{tab:batteryOLO}}
		\begin{tabular}{c c c c c c c c}
			\hline
			$X_R$ & $\beta_R^{k}$ & $\beta_{R|3}^{k}$ & $\beta_{R|(3,6)}^{k}$ & $\beta_{R|(1,3,6)}^{k}$ &
			$\beta_{R|(1,3,4,6)}^{k}$ & $\beta_{R|(1,3,4,6,17)}^{k}$ & Step-wise Sum\\
			\hline
			3 & 0.185 & - & - & - & - & - & 0.185\\
			6 & 0.174 & 0.294 & - & - & - & - & 0.479\\
			1 & 0.102 & 0.165 & 0.279 & - & - & - & 0.758\\
			4 & 0.022 & 0.0350 & 0.051 & 0.087 & - & - & 0.845\\
			17 & 0.024 & 0.027 & 0.036 & 0.067 & 0.079 & - & 0.924\\
			2 & 0.010 & 0.011 & 0.006 & 0.014 & 0.017 & 0.026 & 0.950\\
			\hline
			\multicolumn{8}{c}{Shapley Values}\\
			\hline
			& $X_1$ & $X_2$ & $X_3$ & $X_4$ & $X_6$ & $X_{17}$ &Sum\\
			$\text{Sh}^k_R$ & 0.207 & 0.0132 & 0.333 & 0.053 & 0.302 & 0.044 & 0.950\\
			\hline
			\multicolumn{8}{c}{Select Kernel-ANOVA Indice}\\
			\hline
			$\mathcal{S}_{(1,3)}^k$& $\mathcal{S}_{(1,6)}^k$ & $\mathcal{S}_{(3,4)}^k$& $\mathcal{S}_{(3,6)}^k$ & $\mathcal{S}_{(1,3,6)}^k$ & $\mathcal{S}_{(1,4,6)}^k$ & $\mathcal{S}_{(3,4,6)}^k$ & $\mathcal{S}_{(1,3,4,6)}^k$\\
			0.063 & 0.037 & 0.013 & 0.120 & 0.077 & 0.011 & 0.011& 0.011\\
			\hline
\end{tabular}\end{center}\end{table}

\begin{figure}[t] 
	\center
	\includegraphics[width=\linewidth]{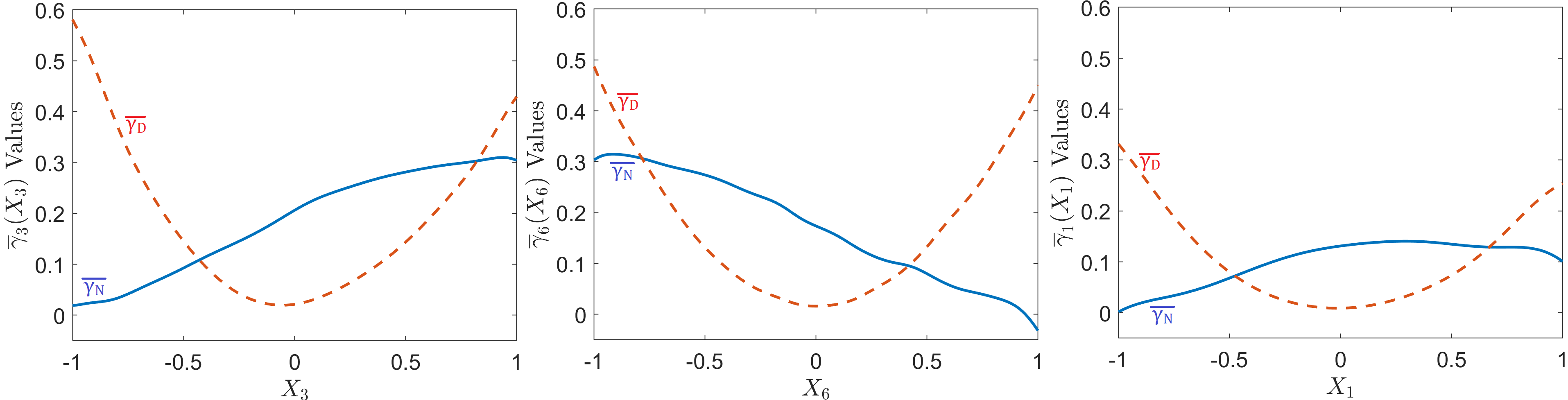}
	\caption{First-order ISFs for (a) $X_3$, (b) $X_6$, and (c) $X_1$.  The blue line gives $\overline{\gamma}_R^{N}$ while $\overline{\gamma}_R^{D}$ is represented by the red line.  The inputs are normalized so $-1$ and $1$ give the minimum and maximum respectively.  The input probability distributions are uniform with constant value and thus not shown.}
	\label{fig:BATISFs}
\end{figure}

\setstretch{1.8}{A first-order sensitivity analysis using the CME procedure is reported in Figure \ref{fig:BUQ1}.  The first-order $\beta_R^k$ indices were used to screen for importance, where inputs $X_1, X_2, X_3,X_4,X_6$ and $X_{17}$ were found to be important where $\beta_R^k>0.005$ was used as threshold of significance. Noting that there are $2^{19}-1$ total unconditional indices, a full-order sensitivity analysis would be untenable. Thus, after determining $\hat{\beta}_R^k=0.950$, the results for only considering the set of significant inputs, $\{X_1, X_2, X_3,X_4,X_6,X_{17}\}$ are presented in Table \ref{tab:batteryOLO}. Since the inputs are independent, the analysis includes the ANOVA effects, $\mathcal{S}_R^k$, that are above the value of $\mathcal{S}_R^k > 0.01$.  The relationship in Eq. \ref{eq:shapcond} can be observed between the conditional and ANOVA indices, for example $\beta_{6|3} = \beta_{6}+\mathcal{S}_{(3,6)}^k=0.294$.  Interestingly, both the Shapley effects and OLS decomposition both agree on the order of importance of the variables being $X_3>X_6>X_1>X_4>X_{17}>X_2$.  The Shapley effects only sum to a value of $0.950$ because they were constructed using only the subset of $\{X_1, X_2, X_3,X_4,X_6,X_{17}\}$.  These results are in reasonable agreement with  \cite{Hadigol2015}, which found the same parameters to be dominant throughout the simulation.  However, the variance-based analysis of \cite{Hadigol2015} was performed at each time-point without output aggregation, which prevents a more detailed comparison. 

The first-order ISFs of the three most significant variables are given in Figure \ref{fig:BATISFs}.  The behavior of the $\overline{\gamma}_R^{N}$ shows the input-output relationship to be marginally heteroscedastic.  For example with inputs $X_1$ and $X_6$ it is observed that the spread of the output data is decreasing as these variables increase, with the inverse relationship for $X_3$.   The behavior of $\overline{\gamma}_R^{D}$ is typical to what was observed in the previous examples, where the conditional and unconditional distributions are most similar when $X_R$ is at the mean value of 0 and grow apart as $X_R$ deviates from the mean.}

This last example demonstrates the relationship given by Eq. \ref{eq:condanovarel} in a scenario where the inputs are independent.  Additionally, the system highlights the significance of operating in the single data set regime, as the computational cost of the simulator is prohibitive for any conditional sampling procedures for a full-order sensitivity analysis.  This concludes the case studies presented in this paper.

\section{Conclusions}\label{sec:conc}
This work proposes significant advances to kernel-based GSA analysis from a single input-output data set.  The utilization of CME allows for the numerical determination of ISFs, which can yield insights into the marginal and cooperative influences of the inputs on the entire output distribution.  Additionally, the ISFs provide a means to compute all of the $\beta_R^k$ indices from a single data set.  Theoretical extensions were made to introduce conditional indices and a new OLS decomposition, which remains valid even when input correlations are present.
These new tools were applied to four illustrative case studies.

The paper also discusses the opportunity for continued research.  Procedures that address the kernel selection process, including hyperparameter determination, for the output kernel are of particular importance.  This goal would likely be aided by detailed study of the asymptotic behavior of the numerical CME estimator with respect to the size of the available data set, building on the works of \cite{Gretton2012,Gretton2012b,Deb2020}. Understanding of these distributions may allow for tailoring the output kernel that maximizes the $\beta^k$-indicators ability to detect input-output dependencies.

\section*{Acknowledgments}
John Barr was supported by the U.S. Department of Energy grant (DE-FG02-02ER15344) and the Princeton Program in Plasma Sciences and Technology (PPST).  Herschel Rabitz acknowledges support from the U.S. Army Research Office grant (W911NF-19-1-0382).

\bibliography{mybibfile}

\end{document}